\documentclass[12pt]{article}
\usepackage{graphicx}

\usepackage{scicite}

\usepackage{times}

\newenvironment{sciabstract}{%
\begin{quote} \bf}
{\end{quote}}

\usepackage{color}

\title{The Emergence of Heterogeneous Scaling in Research Institutions}
\author
{Keith A. Burghardt$^{1\ast}$, Zihao He$^{1}$, Allon G. Percus$^{2,1}$, and Kristina Lerman$^{1}$\\
\\
\normalsize{$^{1}$Information Sciences Institute, University of Southern California, Marina del Rey, USA, 90292}\\
\normalsize{$^{2}$Institute of Mathematical Sciences, Claremont Graduate University, Claremont, USA, 91711}\\
\normalsize{$^\ast$To whom correspondence should be addressed; E-mail:  keithab@isi.edu.}
}

\date{}

\begin{document} 
\baselineskip24pt
\maketitle 


\begin{sciabstract}
Research institutions provide the infrastructure for scientific discovery, yet their role in the production of knowledge is not well characterized. To address this gap, we analyze interactions of researchers within and between institutions from millions of scientific papers. Our analysis reveals that the number of collaborations scales superlinearly with institution size, though at different rates (heterogeneous densification). We also find that the number of institutions scales with the number of researchers as a power law (Heaps' law) and institution sizes approximate Zipf's law.  These patterns can be reproduced by a simple model with three mechanisms: (i) researchers collaborate with friends-of-friends, (ii) new institutions trigger more potential institutions, and (iii) researchers are preferentially hired by large institutions. This model reveals an economy of scale in research: larger institutions grow faster and amplify collaborations. Our work provides a new understanding of emergent behavior in research institutions and how they facilitate innovation.
\end{sciabstract}





Scientific innovation and training require efficient and robust infrastructure, which is provided by research institutions, a category that includes universities, government labs, industrial labs, and national academies~\cite{hicks1996science,Taylor2019}. Despite the long tradition of bibliometric and science of science research~\cite{Fortunato2018}, the focus has only recently shifted from individual scientists~\cite{Sinatra2016,wang2013quantifying} and teams~\cite{Guimera2005,Wuchty2007,milojevic2014principles} to how institutions affect researcher productivity and impact~\cite{Way2019,Deville2014}. Many gaps remain in our understanding of the role of institutions in the production of scientific knowledge, and specifically, how they form, grow, and facilitate scientific collaborations. These questions are important, because collaborations are increasingly prevalent in scientific research~\cite{hicks1996science,Wuchty2007,Guimera2005} and produce more impactful and transformative work~\cite{Wuchty2007,Wu2019}. Collaboration allows scientists to cope with the increasing complexity of knowledge~\cite{jones2009burden} by leveraging the diversity of expertise~\cite{page2019diversity} and perspectives offered by collaborators from different institutions~\cite{Dong2018} and disciplines~\cite{yegros2015does}. 

\begin{figure}[ht]
    \centering
    \includegraphics[width=\columnwidth]{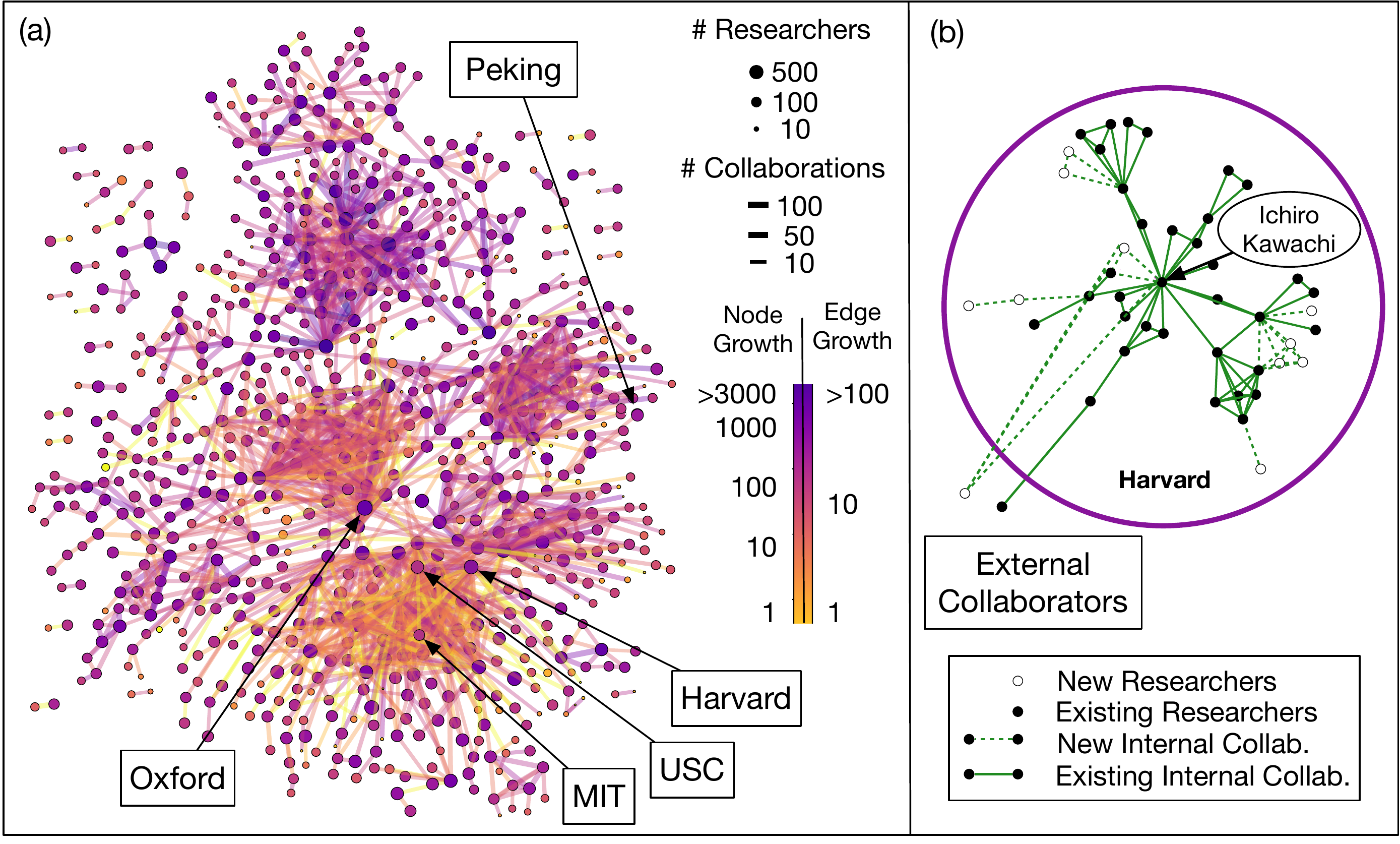}%
    \caption{Collaborations in sociology in 2017.
    (a) External collaborations and institution size. 
    Each node represents a research institution. Institutions with more researchers are represented by larger nodes, and more collaborations are represented by thicker lines (edge weights). Darker nodes represent faster-growing institutions, and darker links represent faster-growing collaborations. Links with fewer than 10 collaborations are removed, as are isolated nodes. (b) The largest connected component of internal collaborations within Harvard University. Each node represents a researcher.
    }
    \label{fig:DataSummary}
\end{figure}

To understand the evolution of research institutions and collaborations, we analyze a large bibliographic database spanning many decades and multiple scientific disciplines. The database contains millions of publications from which the names of authors (collaborators) and their affiliations (research institutions) have been extracted for each paper. Figure~\ref{fig:DataSummary} shows collaboration network between sociology researchers at different institutions, as well as  size of institutions, as of 2017. We see a remarkable diversity of institution size and growth, both in terms of the number of researchers (node growth) and collaborations between institutions (edge growth). Collaborations are clustered, with clear groups of interacting institutions. 
Research collaborations within an institution are equally complex. Figure~\ref{fig:DataSummary}b highlights  the largest connected component of the collaboration network within Harvard. Individual researchers vary widely in the number of collaborators, with new collaborations appearing in clusters.


Analysis of these data reveal strong statistical regularities. We find that 
collaborations scale superlinearly with institution size, i.e., faster than institutions grow, consistent with densification of growing networks~\cite{Leskovec2007,Bhat2016,Lambiotte2016}. However, the scaling law is different for each institution, and as a result, different parts of collaboration network densify at different rates. The scaling laws cannot be explained by growing output (papers), as researcher productivity is roughly constant at each institution. We also find that institutions vary in size by many orders of magnitude, with distribution approximated by Zipf's law \cite{Zipf1949}, while the number of institutions scales sub-linearly with the number of researchers, in agreement with Heaps' law \cite{Lu2010,Simini2019}.  The sublinear scaling implies that, even as more institutions appear, each institution gets larger on average, but this average belies an enormous variance.

We propose a stochastic model that explains how institutions and research collaborations form and grow.
In this model, a researcher appears at each time step and is preferentially hired by larger institutions (e.g., due to their prestige or funding), which leads to the rich-get-richer effect creating Zipf's law. With a small probability, however, a researcher joins a newly appearing institution. The arrival of this new institution then triggers yet more new institutions to form in the future \cite{Tria2014}. Finally, once hired, researchers make connections to other researchers and their collaborators with an independent probability. Despite its simplicity, the model reproduces a range of empirical observations, including the number and size of research institutions, and how pockets of increasingly dense structures form in collaboration networks. Although the first and second step of this model has individually has been studied before, this model is unique by combining these mechanisms to form a cohesive model of collaborations.

These empirical results demonstrate universal emergent patterns in the formation and growth of research institutions and collaborations. Our model demonstrates that new institutions are critical to innovation by providing the triggering mechanism that makes more institutions possible. At the same time, large institutions offer an economy of scale: they grow faster and provide more collaboration opportunities compared to smaller institutions.

\section*{Results}

\begin{figure}
    \centering
    \includegraphics[width=1\columnwidth]{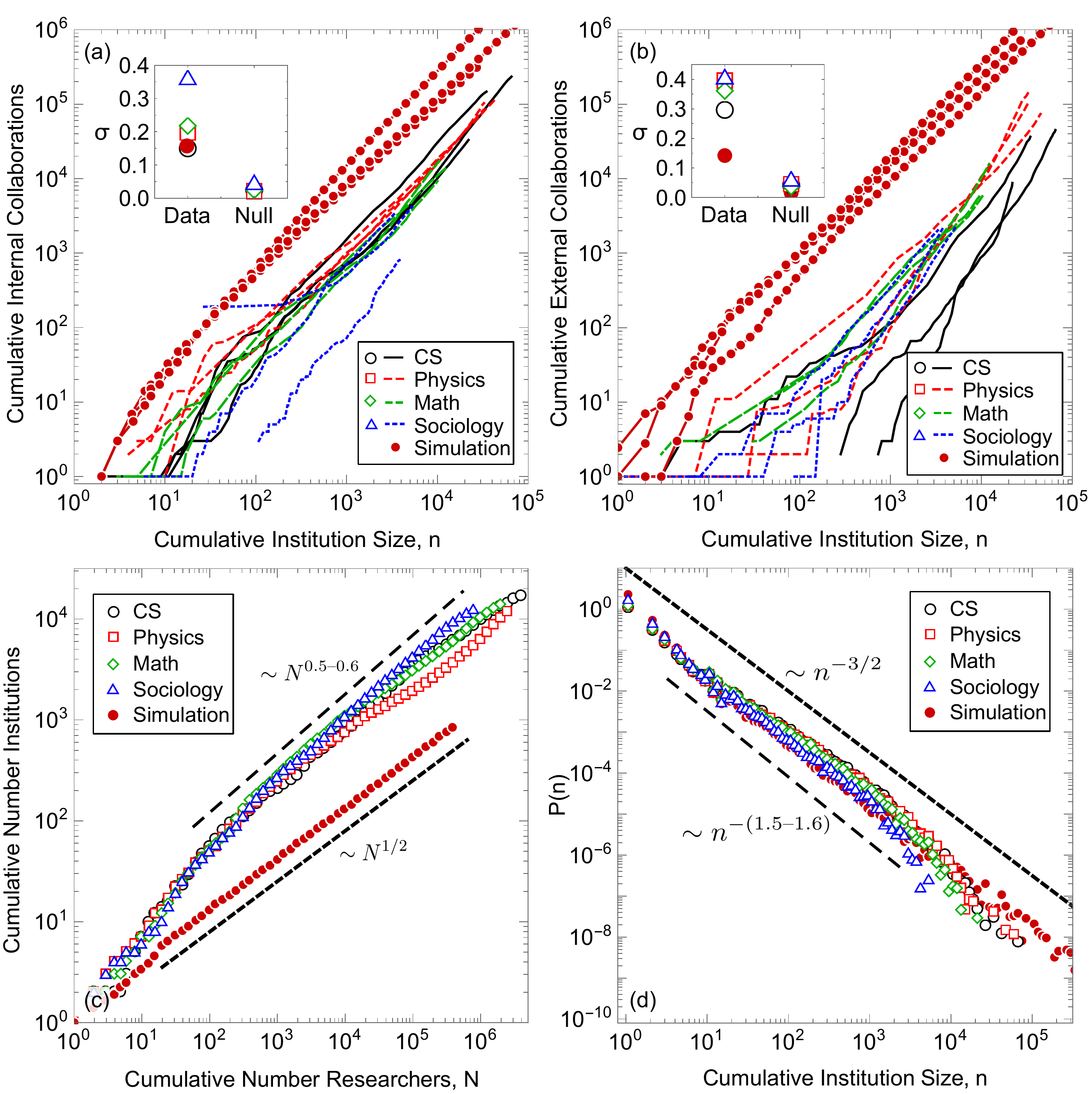}
    \caption{
    Research institution scaling laws. (a) Internal and (b) collaborations versus institution size for three institutions in each field and three simulated  ``institutions''
    . Insets: standard deviation of exponents are much larger than expected by a null model in which differences in exponents are due to statistical fluctuations (see SI Supplementary Note 3 for definition of null model). Standard errors are smaller than plot markers. (c) Number of institutions versus total number of researchers, and (d) the total number of researchers who were ever in each institution as of 2017 (cumulative institution size) for computer science, physics, math, sociology, and a simulation with $\rho=4$, $\nu=2$, $\mu_p=0.6$, and $\sigma_p=0.1$. 
    \label{fig:Scaling}
    }
\end{figure}
As the first step towards characterizing the complexity of institution scaling, we use large longitudinal data to measure how collaborations $c$ scale with institution size $n$. Figure~\ref{fig:Scaling}a--b shows how the number of internal and external collaborations changes over time for several institutions from different disciplines. While each institution follows a scaling law $c \sim n^{\alpha}$ ($R^2$ is consistently around 1, see Supplementary Note 6:  Comparison Between Data and Simulations), the exponents $\alpha$ differ substantially between institutions. 

To show that the scaling exponents of all institutions are different, we create a null model (see SI Supplementary Note 3) in which all institutions follow the same scaling law. In this null model, residuals of each institution's fitted scaling relation are reshuffled and added as noise onto a single scaling relation. Differences between fitted exponents in this model are due to statistical noise rather than different scaling laws. We find that  the variance of the scaling laws across all institutions is much higher than this null model (insets of Fig.~\ref{fig:Scaling}a--b)
. We therefore reject the hypothesis that all the exponents within a field are the same within statistical error. 
We explore the dependence of scaling on institution size, $n$, in Supplementary Note 6:  Comparison Between Data and Simulations, and find the scaling exponents are superlinear (approximately 1.2 on average) and independent of $n$. 
This implies that parts of the collaboration network densify at different rates, which are intrinsic to each institution.

We find weak evidence that higher scaling exponents correspond to institutions with greater success. In physics, the Spearman rank correlation, $s$, between mean paper impact after five years and internal collaboration scaling exponents is 0.09 (borderline significant, p-value$=0.06$) and for external collaboration is 0.27 (p-value $<10^{-5}$). Similarly, in sociology, the correlation is 0.19 (p-value$=0.03$) between impact and internal collaboration exponents, and the same correlation value is found for external collaboration exponents. For all other fields, however, the correlations are not statistically significant (p-value$\ge 0.20$). Impact, a proxy of institution research quality, cannot fully explain why collaborations grow faster in some institutions and not others, but can give some insight into reasons for this diversity. 


The superlinear scaling of collaborations cannot be explained by the higher productivity of researchers at larger institutions. When we look at institution output, i.e., the cumulative number of papers published by researchers affiliated with that institution at a given year, the scaling exponents of output are centered around 1.0 (see SI Supplementary Note 4: Scaling of Output). This suggests that paper output per researcher is approximately independent of institution size. Instead, we find that average team size per institution  increases with institution size (see SI Supplementary Note 5: Scaling of Team Size), which explains the scaling of collaborations. 



\begin{table}[t]
    \centering
{
    \begin{tabular}{c|c|c}\hline
        \vspace{-10pt}
    &&\\
    \textit{Discipline} & \textit{Heaps' Law Exponent} & \textit{Zipf's Law Exponent} \\\hline
    Comp. Sci. & $0.554 \pm 0.004$&$-1.470 \pm 0.005$ \\
    Physics &$0.501\pm 0.007$ & $-1.474 \pm 0.006$\\
    Math &$0.549 \pm 0.008$ & $-1.516\pm 0.006$\\
    Sociology &$0.622 \pm 0.005$ & $-1.603 \pm 0.009$\\
    Simulation &$0.5$ & $-1.5$\\\hline
    \end{tabular}
    }
    \caption{Zipf's Law and Heaps' Law Exponents}
    \label{tab:scaling}
\end{table}

We also find that the number of institutions grows sublinearly with the number of researchers (Fig.~\ref{fig:Scaling}c): as new researchers start their careers, new institutions eventually form. The number of institutions follows Heaps' law \cite{Tria2014}.
The distribution of institution sizes (as of 2017), on the other hand, follows Zipf's law (Fig.~\ref{fig:Scaling}d), similar to 
the observed heavy-tailed distribution of city sizes~\cite{Zipf1949,Batty2006}. 
Exact scaling law values for each field can be found in Table~\ref{tab:scaling}, where scaling laws are calculated for the number of researchers, $N$, greater than twenty 
and institution size, $n$, greater than ten.

\subsection*{A Model of Institution Growth}

We now describe a stochastic growth model of institution formation that elucidates how institutions and collaborations jointly grow. We model institution formation and growth with a P\'olya's urn-like set of mechanisms described in \cite{Tria2014}, and we model the growth of collaborations with a network densification mechanism~\cite{Bhat2016,Lambiotte2016}. Unlike existing models of network densification~\cite{Leskovec2007,Bhat2016,Lambiotte2016},  however, our model reproduces the heterogeneous densification of internal and external collaborations, and the non-trivial growth structure on institutions.

\begin{figure}[tb!]
    \centering
    \includegraphics[width=1\columnwidth]{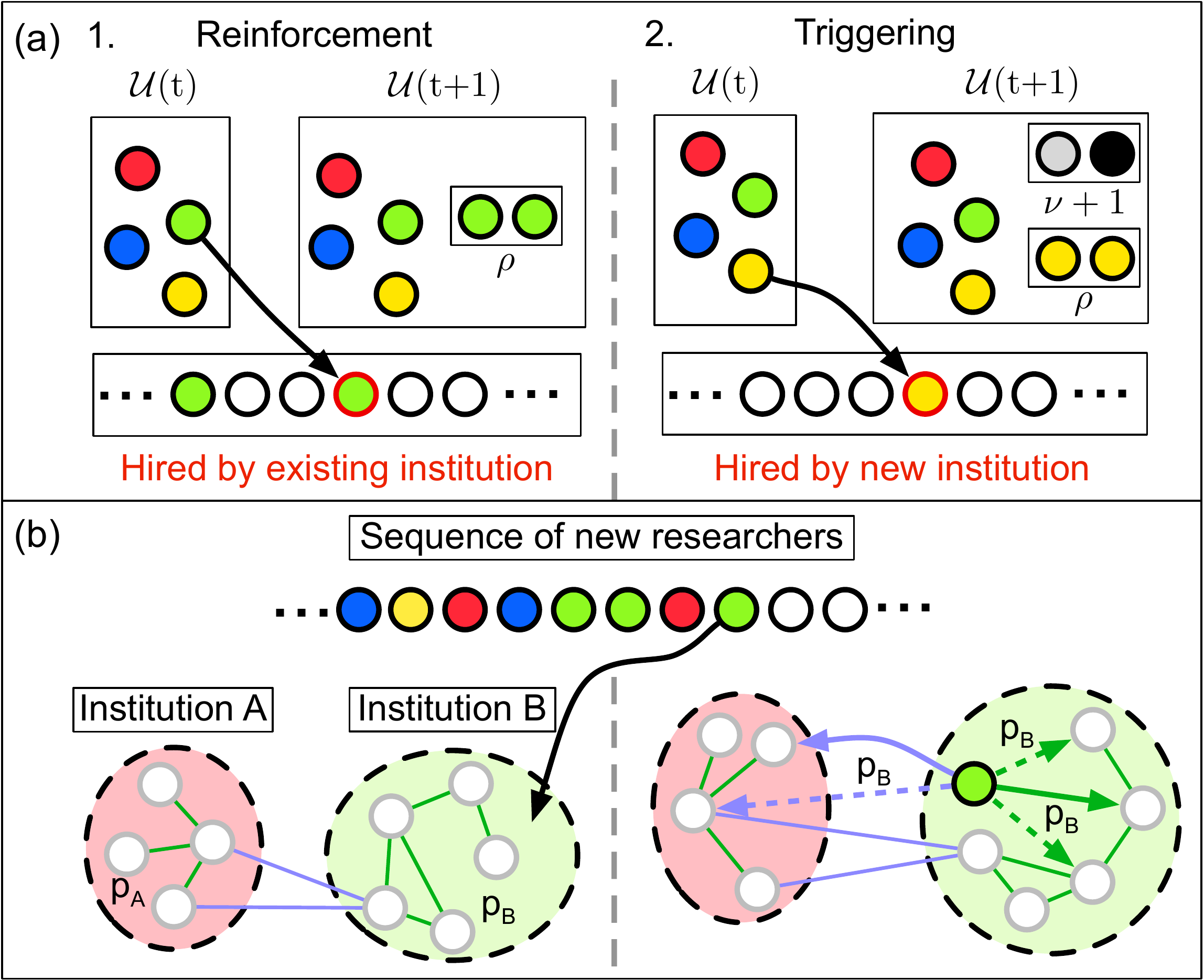}
    \caption{Schematic of our institution growth model. (a) New researchers are hired by an institution following a P\'olya's urn-like model \cite{Tria2014}. In this model, a new researcher is hired by an institution, denoted by a colored ball, picked uniformly at random from an urn. A new institution, where no researcher has been hired before, triggers $\nu+1$ new colors to enter the urn, increasing the likelihood of more new institutions to hire a researcher. Both new and old institutions experience reinforcement, where $\rho$ balls of the same color enter the urn. This creates a rich-get-richer effect where large institutions are more likely to hire a new researcher. (b) Each institution is composed of both internal collaborators (green lines) and external collaborators (purple lines). Once a researcher is hired, they choose one random internal and one random external collaborator. New collaborations are formed independently with probability $p_A$, if hired by institution A, and $p_B$ if hired by institution B. These new connections form triangles. 
    }
    \label{fig:ModelSchematic}
\end{figure}

We imagine an urn containing balls of different colors, with each color representing a different institution, as shown in  Fig.~\ref{fig:ModelSchematic}a. Balls are picked with replacement, each pick representing a newly hired researcher, and the ball color is recorded in a sequence to represent what institution hires the researcher. Afterwards $\rho$ balls of the same color are added to the urn to represent the additional resources and prestige given to a larger institution, known as ``reinforcement'' (left panel of Fig.~\ref{fig:ModelSchematic}a)~\cite{Tria2014}. If a previously not seen color is chosen, then $\nu+1$ uniquely-colored balls are placed into the urn, a step known as ``triggering'' (right panel of Fig.~\ref{fig:ModelSchematic}a)~\cite{Tria2014}. The new colors represent institutions that are able to form because of the existence of a new institution. For example, UC Davis was spun out of UC Berkeley, and USC Institute for Creative Technology was spun out of USC Information Sciences Institute, which itself was founded by researchers from the Rand Corporation. This model predicts Heaps' law with scaling relation $\sim N^{\nu/\rho}$ and Zipf's law with scaling relation $\sim n^{-(1+\nu/\rho)}$ \cite{Tria2014}. In our simulations, we chose $\rho=4$ and $\nu=2$, which agrees well with the data shown in Fig.~\ref{fig:Scaling}. 

Next, we explain heterogeneous and superlinear scaling of collaborations through a model of network densification. Building on the work of  \cite{Lambiotte2016,Bhat2016}, we have each new researcher, represented as a node, connect to a random researcher within the same institution, as well as an external researcher picked uniformly at random (left panel of Fig.~\ref{fig:ModelSchematic}b). New collaborators are then chosen independently from neighbors of neighbors with probability $p_{i}$, where $p_{i}$ is unique to each researcher's institution (right panel of Fig.~\ref{fig:ModelSchematic}b). 
We let $p_{i}$ be a Gaussian distributed random variable with mean, $\mu= 0.6$, and standard deviation,  $\sigma_\mu=0.1$ and truncated between 0 and 1. We show separately that $p_{i}$ directly controls the heterogeneity we observe in internal collaboration scaling, but the heterogeneity in  external collaboration scaling is an emergent outcome of this model \cite{BurghardtPRE2021}.

To summarize, our model has four parameters: $\rho=4$, $\nu=2$, $\mu_p=0.6$, and $\sigma_p=0.1$. This model reproduces Heaps' and Zipf's laws (Fig.~\ref{fig:Scaling}c--d and Table~\ref{tab:scaling}) and the heterogeneous scaling of internal and external collaborations shown in Fig.~\ref{fig:Scaling}a--b. While other plausible mechanisms for Zipf's law \cite{Gibrat1931,Eeckhout2004,Axtell2001}, Heaps' law  \cite{Simini2019}, or densification \cite{Leskovec2007} exist, the current model describes these patterns in a cohesive framework and explains the heterogeneous scaling we discover in the data. While this heterogeneity is built into our internal scaling laws, the external scaling heterogeneity is a uniquely emergent property within the model \cite{BurghardtPRE2021}.

The model also reproduces qualitative trends of cross-sectional analysis. Specifically,  the scaling exponents of internal collaborations produced by the model when measured at a specific point in time, i.e., in cross-sectional setting, vary in time and are larger than scaling exponents of external collaborations and decrease over time (SI Supplementary Note 6: Comparison between Data and Simulations), unlike what we see in data (SI Supplementary Figure 3). 
These results are robust to stochastic variations of the densification mechanism (SI Supplementary Note 7: Simulations of Alternative Mechanisms). As a final comparison with data, we compared the growth of institutions and the ways links form to the model mechanisms and found broad agreement \cite{BurghardtPRE2021}.

\section*{Discussion}

We identify strong statistical regularities in the growth of research institutions. The number of collaborations increases superlinearly with institution size, i.e., faster than institutions grow in size, though the scaling is heterogeneous, with a different exponent for each institution. The superscaling is not explained by the increased productivity of researchers at larger institutions---the number of papers per researcher is roughly independent of institution size. Instead, the growing collaborations are associated with bigger teams at larger institutions. The diversity in collaboration scaling exponents is partly explained by variations in institution impact. Institutions with higher impact papers also tend to have a larger scaling exponent. This provides evidence that a higher collaboration scaling exponent allows for collaborations to form more easily, and that in turn creates higher-impact papers. Further analysis is needed to test this hypothesis in the future.

When these observations are incorporated into a minimal stochastic model of institution growth, we are able to reproduce the surprising regularity of research institution formation, growth and the heterogenous densification of collaboration networks. 
These findings support the idea that academic environments differ in their ability to bolster researcher productivity and prominence~\cite{Way2019}, and also demonstrate that institution size and ability to facilitate collaborations as a potential factor explaining differences in academic environments. Additional research is also needed to identify other factors that contribute to an institution's success.

\section*{Methods \& Materials}
\subsection*{Data}
We use bibliographic data from Microsoft Academic Graph (MAG), from which researcher names (authors), their institutional affiliation, and references made to other papers have been extracted~\cite{sinha2015overview,herrmannova2016analysis}. MAG data has disambiguated institutions and authors for each paper, allowing us to consider all authors with the same unique identifier to be the same researcher, and similarly for each institution. The MAG data only records one affiliation per researcher per paper, even when the researcher may have multiple affiliations in a given paper. Such cases are rare, especially among older papers \cite{Hottenrott2017}, and thus unlikely to affect our results. 
We focus on four fields of study: computer science, physics, math and sociology. 
After data cleaning, we have almost ten million papers published between 1800 and 2018 (see SI Supplementary Note 1). Our computer science data includes early research in topics relating to computers, including electrical engineering, and therefore stretches back to before 1900.

We define \textit{institution size} in a given year as the number of authors who have been ever been affiliated with that institution up until that year. \textit{Collaborations} are defined as two researchers who have co-authored a paper up until that year. We distinguish between \emph{internal collaborations} (co-authors at the same institution) and \emph{external collaborations} (co-authors affiliated with different institutions). Finally, to understand the relation between collaborations and institution size, we define \emph{output} as the cumulative number of papers from researchers affiliated with an institution in a particular year.

\subsection*{Analysis}
We use cumulative statistics to reduce statistical variations and to better compare to a stochastic growth model of institution formation. To check the robustness of results, we compare to an alternate yearly definition of institution size and collaborations (see SI Supplementary Note 2: Cumulative versus Yearly Measures). We find all qualitative results are the same, in part because both definitions are highly correlated. 

We present scaling results for longitudinal analysis, which tracks how collaborations evolve as individual institutions grow~\cite{depersin2018global,Keuschnigg2019,Ribeiro2020}. This contrasts to cross-sectional analysis applied in previous work on city scaling \cite{Bettencourt2007,Bettencourt2013} and institution scaling \cite{Taylor2019}, which measures collaborations as a function of size of all institutions at a given point in time. We find that cross-sectional analysis identifies scaling laws that are not representative of the growth of most institutions (see Supplementary Note 7: Robustness Check of Simulations), and while simulations and empirical data give  scaling exponents that are fairly constant in time for each institution, cross-sectional scaling exponents vary in time for both data and simulation. 
For these reasons, we focus on longitudinal scaling analysis in this paper, 
although scaling laws derived by either analysis method strongly relate to each other \cite{Ribeiro2020,Bettencourt2020}.






\section{Supporting Information}

\subsection*{Supplementary Note 1: Data}
\label{si-data}
We use bibliographic data from Microsoft Academic Graph (MAG)\footnote{{https://www.microsoft.com/en-us/research/project/microsoft-academic-graph/}}, from which researcher names (authors), their institutional affiliations, and references made to other papers have been extracted~\cite{sinha2015overview,herrmannova2016analysis}. 
MAG data has disambiguated institutions and authors for each paper, allowing us to consider all authors with the same unique identifier to be the same researcher, and similarly for each institution. The data only records one affiliation per researcher per paper, even when researchers may have multiple affiliations within a given paper. However, multiple affiliations are rare~\cite{Hottenrott2017} and can be safely ignored.

The MAG data enables us to measure institution size (the number of published authors affiliated with the institution), productivity (number of papers written), and collaborations (co-authors of the same paper), both within and among institutions. We gather data from papers published in four fields of study between 1800 and 2018: computer science (14,666,855 papers), physics (8,428,923 papers), math (6,192,706 papers), and sociology (4,407,288 papers). Because the metadata for MAG are extracted automatically, many papers have some missing values among extracted names, references, institution, or year published. 
As part of the data cleaning process, we remove papers with missing fields, and also papers with more than 25 authors. These many-authored papers only represent $0.70\%$ of all physics papers, and $<0.036\%$ of papers in other fields but are removed because they may be 
too large to constitute a meaningful collaboration between any individuals. This leaves 3,916,332 computer science papers, 
2,494,000 physics papers, 
2,370,712 math papers, 
and 1,115,841 sociology 
papers. 

\begin{figure}[thb!]
    \centering
    \includegraphics[width=0.6\textwidth]{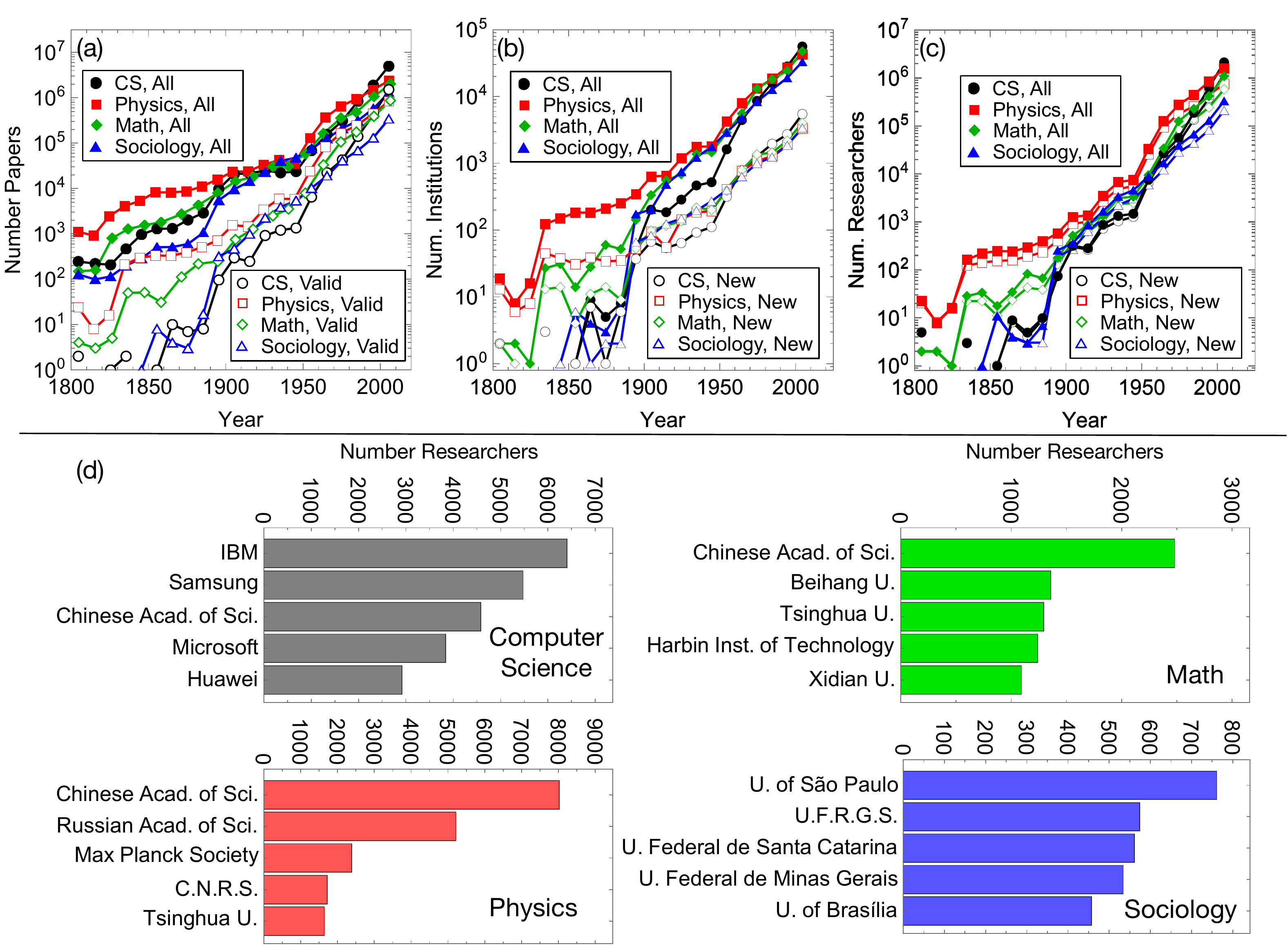}
    \caption{Growth of academic disciplines. (a) Number of papers per decade and number of valid papers that contain author, references, institution, and year. (b) Number of new institutions over time. (c) number of new researchers over time. (d) Largest institutions in 2017. C.N.R.S. stands for Centre national de la recherche and U.F.R.G.S. stands for Universidade Federal do Rio Grande do Sul. scientifique}
    \label{fig:NumResInstStatsPerYear}
\end{figure}

Supplementary Figure~\ref{fig:NumResInstStatsPerYear} shows the descriptive statistics of the data, including the growth of the number of researchers, institutions, and papers published in the four disciplines, and the five largest institutions in each field. Notably, while the largest Physics, Sociology, and Math institutions are universities, the largest computer science institutions are often companies. Figure 2 in the main text demonstrates that institution sizes are broadly distributed with many smaller than 10 researchers, and some larger than $10^3$. While Supplementary Figure~\ref{fig:NumResInstStatsPerYear} shows that the largest institutions are intuitive, such as Harvard, we separately check the quality of the data for small institutions. We randomly sampled 44 institutions in each field with fewer than 10 researchers as of 2017. We observe that they tend to be for-profit colleges, community colleges, and institutions without a formal department in the field of interest (e.g., an engineering school with papers in sociology). That said, we see the journals they publish in tend to be well-aligned with the field, therefore the small institutions were not associated with a particular field by mistake. While these are qualitative checks, they nonetheless show that the data and institutions found are reasonable. 

\begin{figure}[thb!]
    \centering
    \includegraphics[width=0.3\textwidth]{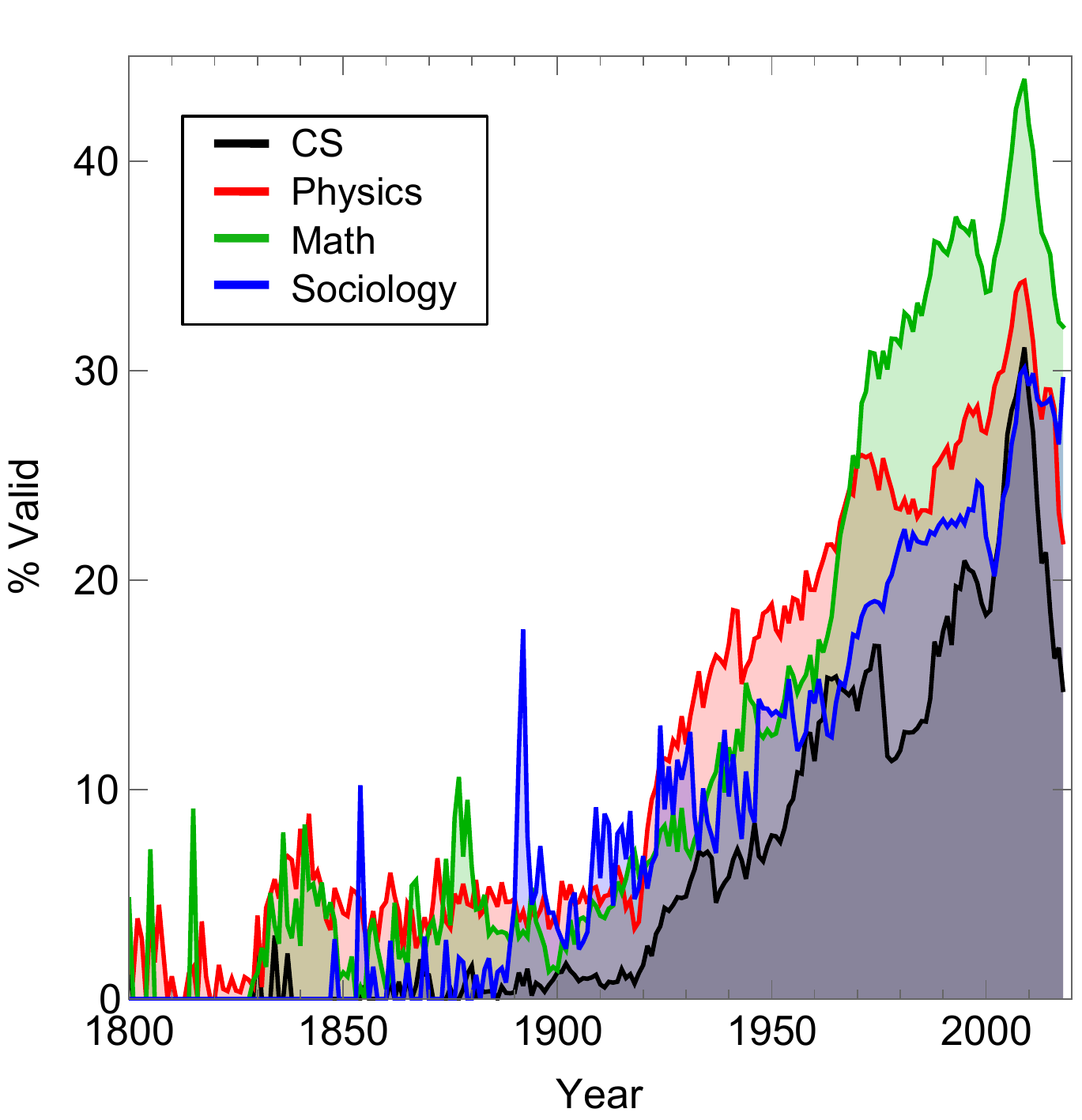}
    \caption{Percentage of valid papers each year. Valid papers are defined as those with names, years, and affiliations for the authors. We find newer papers, especially after 1900, are more likely to be valid.}
    \label{fig:pctvalid}
\end{figure}

\begin{figure}[thb!]
    \centering
    \includegraphics[width=0.8\textwidth]{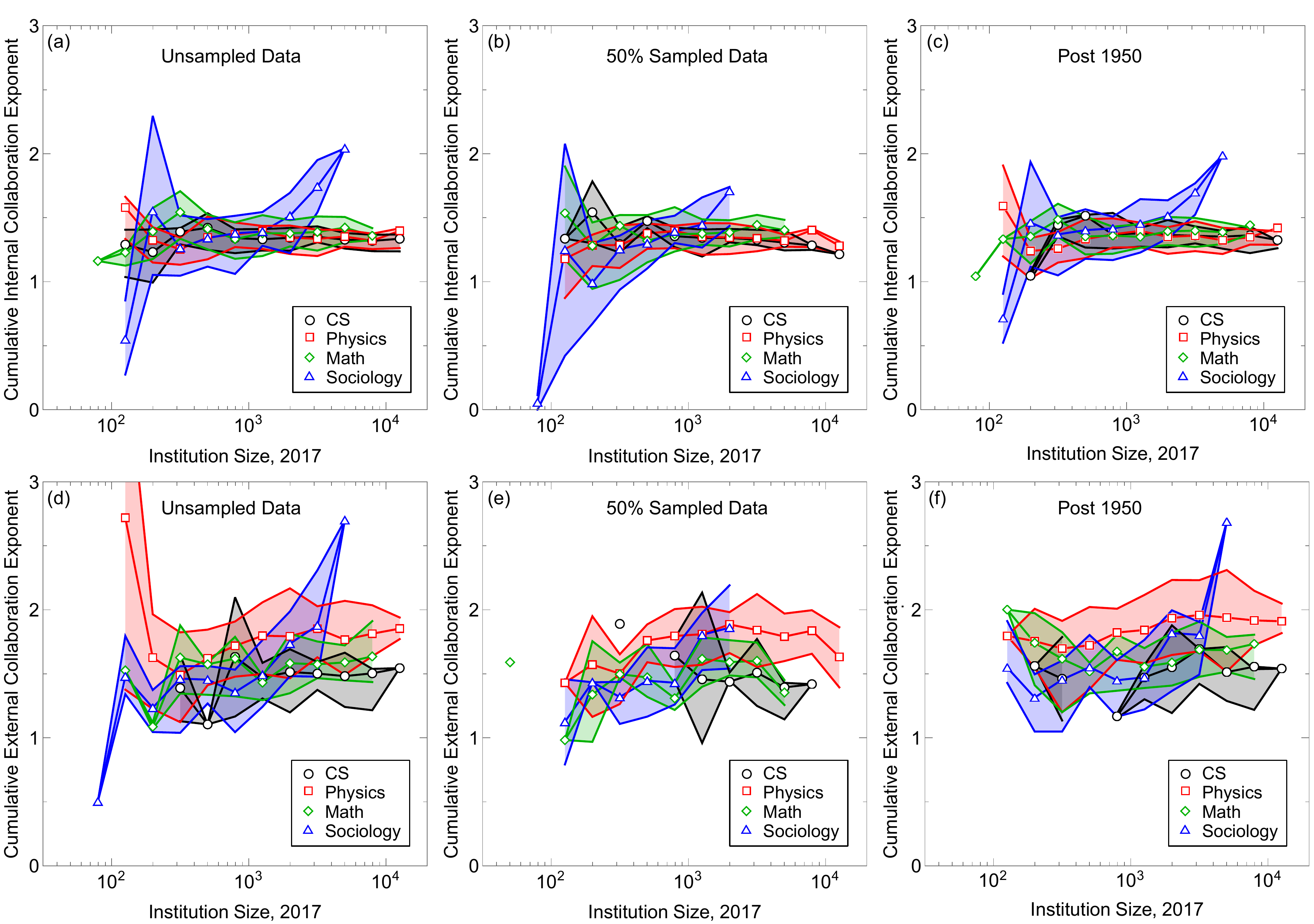}
    \caption{
    Collaboration scaling for unsampled data and data in which 50\% of all papers are removed at random. (a--b) Cumulative internal collaborators vs.\ final 
    institution size, $n$, for (a) unsampled data, (b) 50\% sampled data, and (c) data from after 1950 (which has better sampling). 
    (d--f) External collaboration scaling exponents versus final institution size for (cd unsampled data, (e) 50\% sampled data, and (f) data from after 1950.}
    \label{fig:Sampled}
\end{figure}

We also analyze the quality of MAG's data over time in Fig.~\ref{fig:pctvalid}. This figure shows the percentage of papers considered valid (containing year, author, and affiliation). We find surprisingly few papers are considered valid before 1900, while even when the data quality is highest around 2000, the minority of papers are considered valid. This demonstrates bias in data sampling, and is a potential limitation of our work. 
Nonetheless, we show in Supplementary Figure~\ref{fig:CrossInstituteProduction} \& ~\ref{fig:ProductionScaling} that the number of papers per author is roughly independent with institution size. If the sampling bias had an increasing preference towards, e.g., large institutions, then this finding would not have held. Furthermore, Supplementary Figure~\ref{fig:Sampled} shows that removing half of all data, or removing data pre-1950 (where the percent of valid data is low), does not substantially affect the scaling relations. In that plot, we show the scaling relations versus institution size in 2017 for all institutions studied. Data is qualitatively and quantitatively the same despite the significant drop in the number of papers studied. These robustness tests suggest that undersampling does not affect our overall conclusions. 


Finally, because we include all research output, including journal papers and patents, we check the robustness of our findings when just including the standards of academic research: journals and conference proceedings. In Table~\ref{tab:journal}, we find that between 57\%-90\% of all documents are in these two categories, with 90\% in sociology and only 57\% in computer science, presumably because research output in that field is often patents. In Supplementary Figures~\ref{fig:JournalGrowth} \& \ref{fig:JournalDensification}, however, we show that our major findings are qualitatively unchanged. We see a strong Heaps' law and Zipf's law that looks very similar to the main text (Fig.~\ref{fig:JournalGrowth}). Next, we find in Fig.~\ref{fig:JournalDensification}, that collaboration scaling laws are virtually unchanged from the main text figures, thus our results are robust to different forms of data cleaning.

\begin{table}[tbh!]
\centering
\caption{Journals, conferences and other papers.}
\begin{tabular}{l|cccc}
Data & Journal \& Conference &Other & Total & \% J\&C \\\hline\vspace{-9pt}
&&&\\
CS & 1571485 & 1172424& 2743909 &57\%\\
Physics  & 1953468 & 287740 & 2241208 & 87\%\\
Math & 1567928 & 401392 & 1969320 &80\%\\
Sociology & 913549 & 101146 & 1014695&90\%\\
\end{tabular}\label{tab:journal}
\end{table}

\begin{figure}[thb!]
    \centering
    \includegraphics[width=0.8\textwidth]{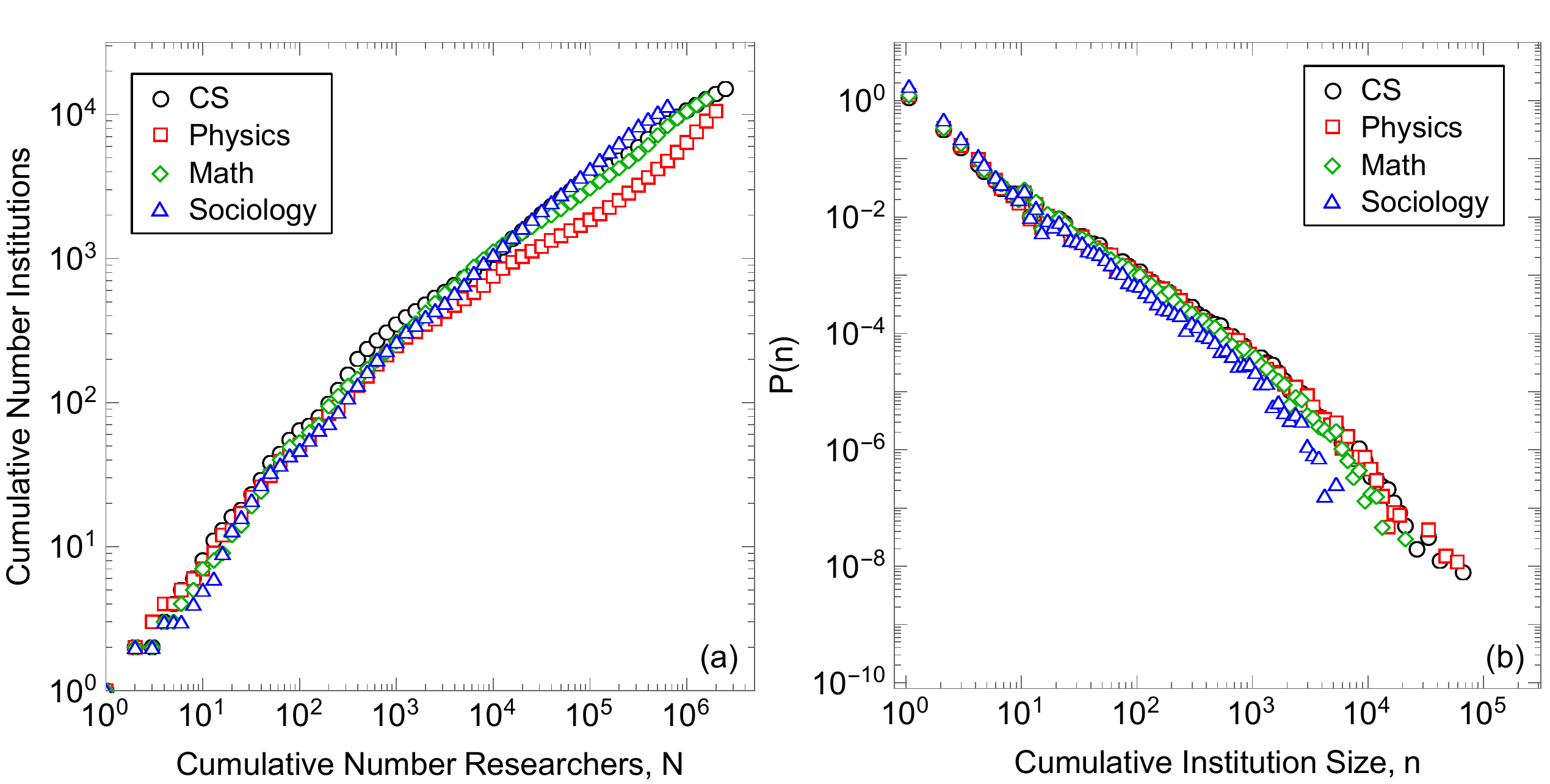}
    \caption{Heaps' law and Zipf's law based on documents labeled as journals or conference proceedings.
    (a) Cumulative number of institutions versus number of researchers  for computer science, physics, math, and sociology. (b) Institution size distribution in 2017.}
    \label{fig:JournalGrowth}
\end{figure}

\begin{figure}[thb!]
    \centering
    \includegraphics[width=0.8\textwidth]{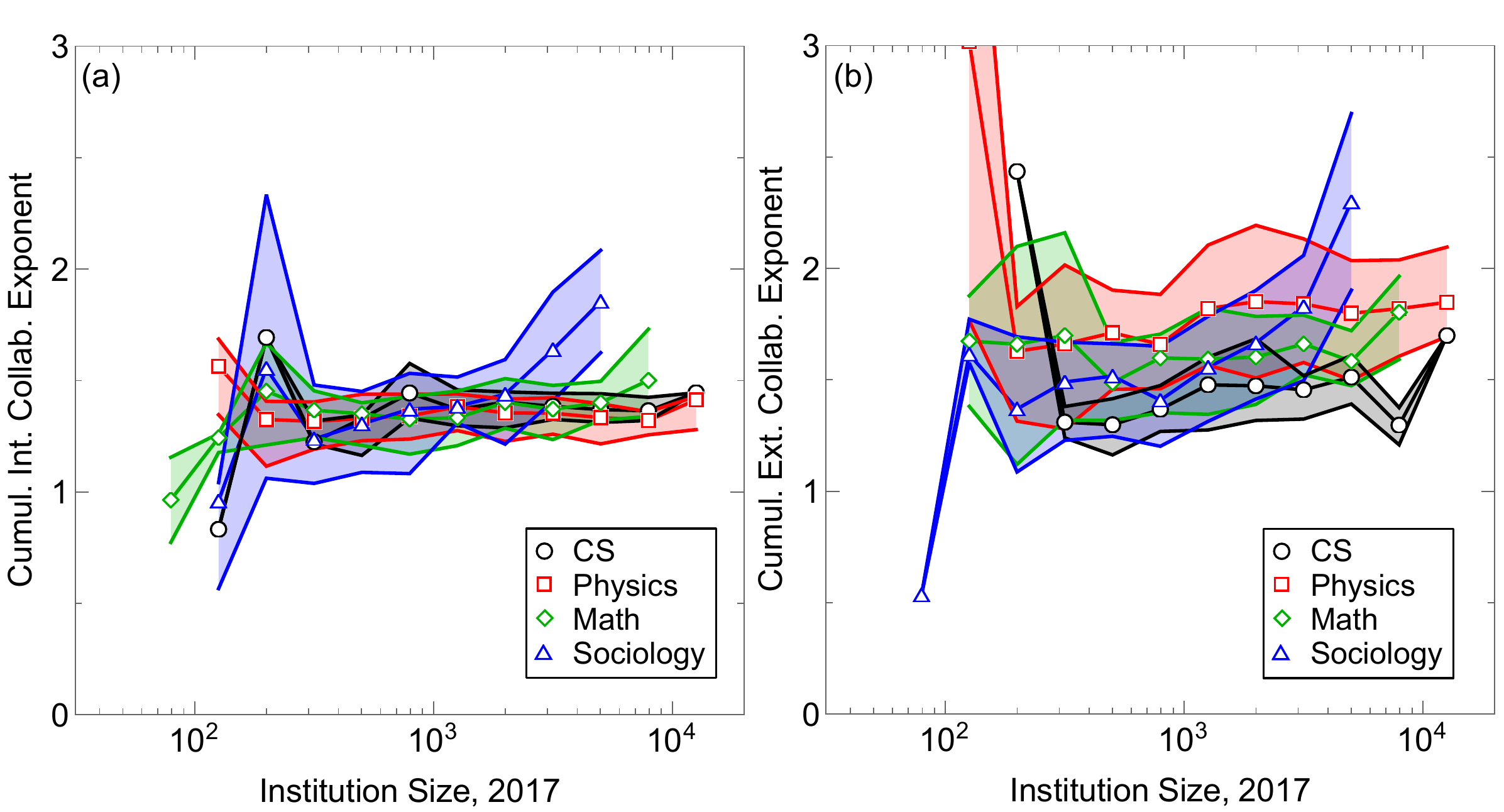}
    \caption{
    Collaboration scaling for authors who coauthor documents labeled as journals or conference proceedings. (a) Cumulative internal collaborators vs.\ final 
    institution size, $n$, (b) external collaboration scaling exponents versus final institution size.}
    \label{fig:JournalDensification}
\end{figure}
\subsection*{Supplementary Note 2: Cumulative versus Yearly Measures}
\label{si-cumulative}
While the main text measured the cumulative size of institutions and collaborations, 
the findings are qualitatively the same if the growth of research institutions was measured on a year-to-year basis. Institution size is therefore the number of active authors affiliated with that institution who published in that particular year. Collaborations were similarly based on papers published that year, etc. 

Figure~\ref{fig:NumResInstStatsPerYear} shows the growth of four academic disciplines, including (a) the number of published papers, (b) the number of  institutions and (b) the number of researchers each year all increase exponentially, regardless of whether these are measured on the cumulative (all) or year-to-year basis. 

\begin{figure}[thb!]
    \centering
    \includegraphics[width=0.9\textwidth]{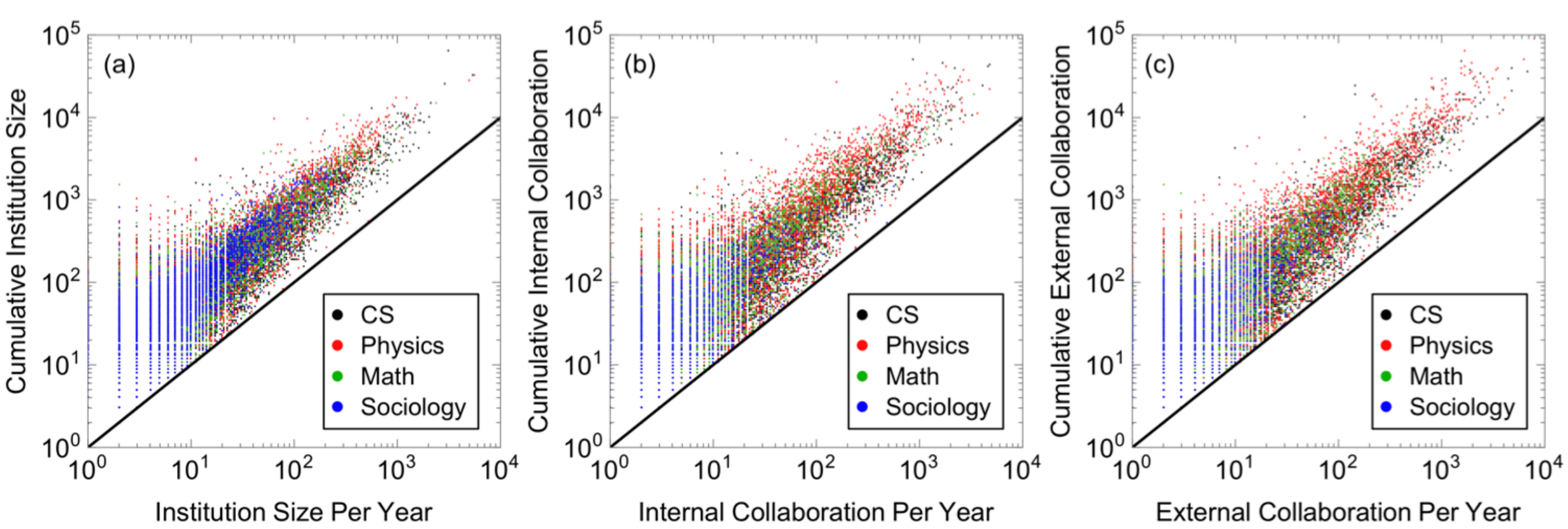}
    \caption{Cumulative versus year-to-year data. (a) Cumulative versus year-to-year institution size, (b) cumulative versus year-to-year internal collaborations, and (c) cumulative versus year-to-year external collaborations.}
    \label{fig:CumulVsYearlyCorrel}
\end{figure}

\begin{table}[tbh!]
\centering
\caption{Cumulative Vs. year-to-year Spearman Correlations}
\begin{tabular}{l|ccc}
Data & Size &Internal Collab. & External Collab. \\\hline\vspace{-9pt}
&&&\\
CS  & 0.85 & 0.83 & 0.83\\
Physics & 0.85 & 0.84 & 0.84 \\
Math & 0.84 & 0.82 & 0.82 \\
Sociology & 0.81 & 0.71 & 0.71\\
\end{tabular}\label{tab:correl}
\end{table}

\begin{figure}[thb!]
    \centering
    \includegraphics[width=0.6\textwidth]{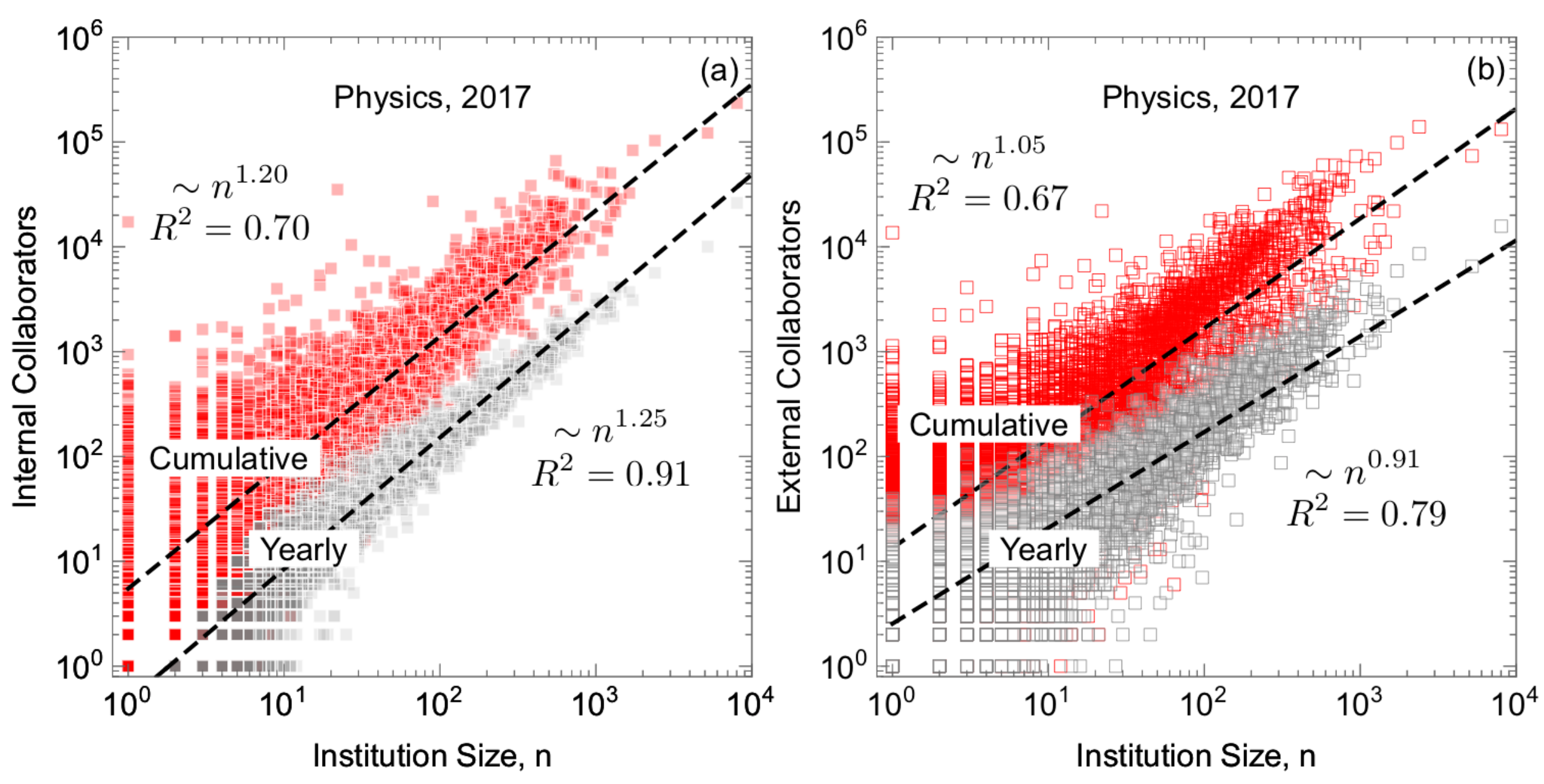}
    \caption{
    Cross-sectional analysis of the scaling of collaborations.
    (a) Cumulative internal collaborators and (b) cumulative external collaborators versus 
    institution size (as of 2017), $n$, for physics. 
    }
    \label{fig:YearlyVsCumScalingOverTime}
\end{figure}

Figure~\ref{fig:CumulVsYearlyCorrel} further demonstrates the robustness of our results, regardless of how they are measured. This figure shows that the cumulative institution size, cumulative number of internal and external collaborations are well correlated with their year-to-year values. The correlations are in Table~\ref{tab:correl}, where Spearman correlations are 0.7--0.8 or higher. Comparison between yearly and cumulative results can also be seen in Fig.~\ref{fig:YearlyVsCumScalingOverTime} where we show cross-sectional collaboration scaling for researchers active in 2017 as well as cumulative collaboration scaling for cumulative institution size. 

\begin{figure}[thb!]
    \centering
    \includegraphics[width=0.6\textwidth]{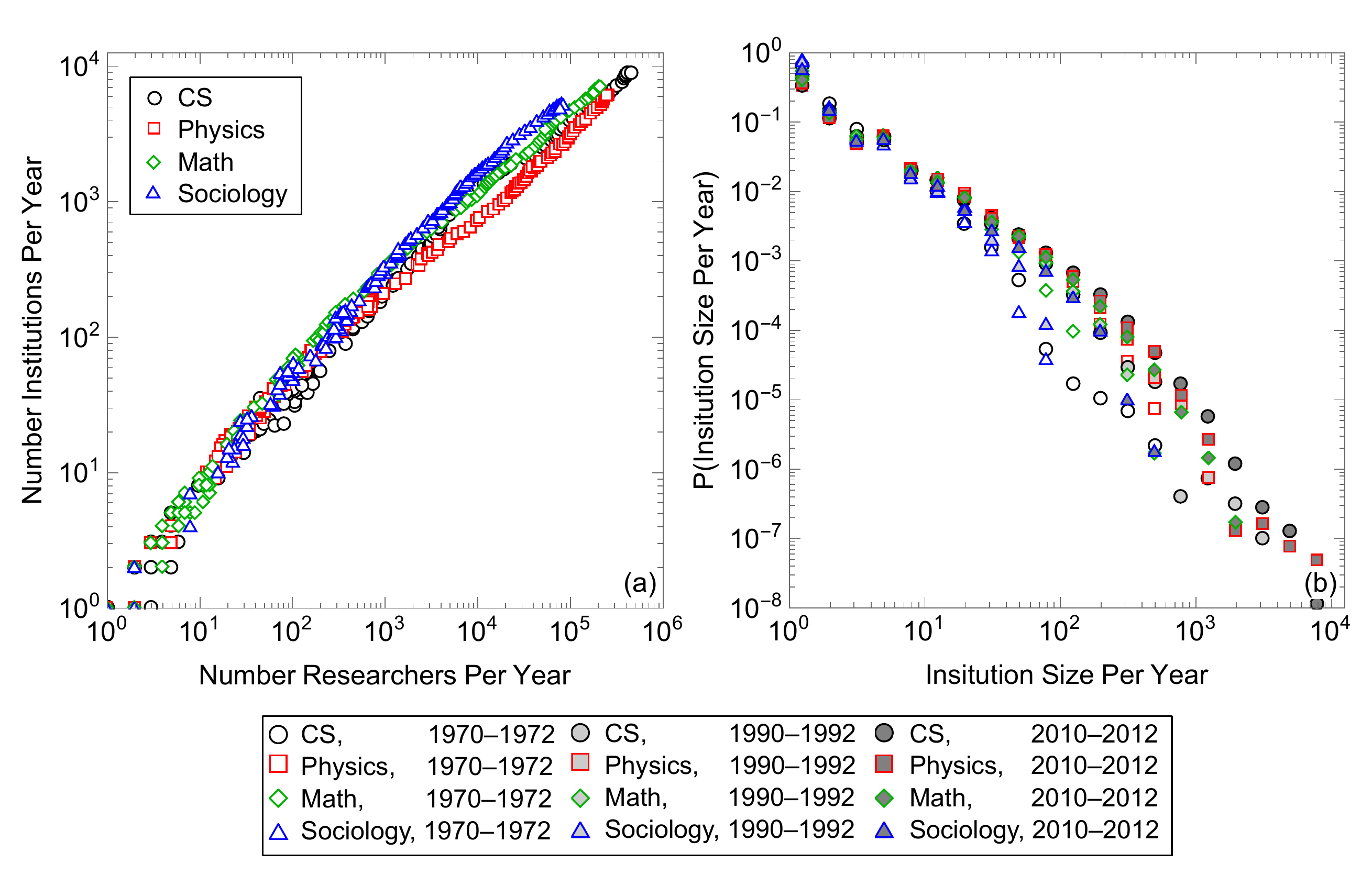}
    \caption{Institution growth statistics by year. (a) Number of institutions versus number of researchers each year for computer science, physics, math, and sociology. (b) Institution size distribution for 1970--1972, 1990--1992, and 2010--2012 for the same fields.}
    \label{fig:PerYearGrowth}
    \end{figure}

Figure~\ref{fig:PerYearGrowth} reproduces Fig. 4 in the main text, except we calculate the yearly number of researchers, institutions, and institution size. Figure~\ref{fig:PerYearGrowth}a shows the number of institutions in a given year versus the number of researchers in a given year. We see, much like in the main text, a sub-linear scaling between the number of institutions and researchers. Figure~\ref{fig:PerYearGrowth}b shows the institution size distribution. Importantly, the institution size distribution might change over time, therefore we plotted the institution size per year for 1970--1972, 1990--1992, and 2010--2012, and found the distribution was extremely stable in time and across fields.

\begin{figure}
    \centering
    \includegraphics[width=0.5\textwidth]{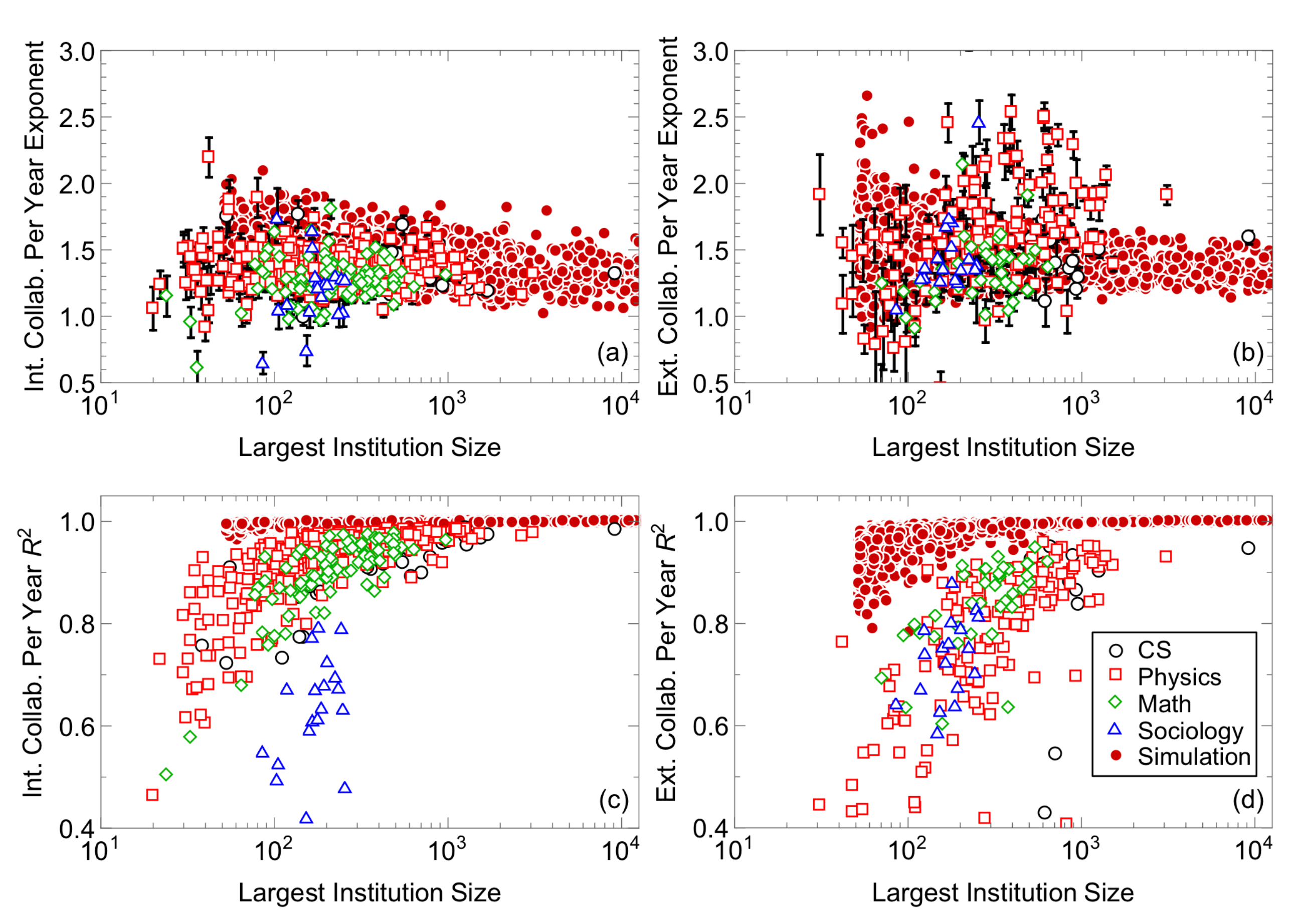}
    \caption{Parameter fits versus 
    institution size per year. (a) Internal and (b) external collaboration scaling exponents versus institution size for collaborations per year. $R^2$ is lower for smaller institutions but quickly approaches 1.0 for (c) internal and (d) external collaboration per year. Compare to Fig. 3 in the main text.}
    \label{fig:PerYearBetaR2VsN}
\end{figure}

Figure~\ref{fig:PerYearBetaR2VsN} shows the year-to-year collaboration scaling exponents versus the largest institution size, as well as the quality of the linear model fits ($R^2$). We see, much like the main text, a large variance in the exponent values, but that they do not significantly change with institution size. That said, the scaling law $R^2$ is higher for large institutions in agreement with the expectation that the scaling law works best in the large-$n$ limit of institution sizes. This is also similar to what was found in the main text for cumulative sizes.

\subsection*{Supplementary Note 3: Homogeneous Densification (Null) Model}
To understand whether the observed heterogeneity in longitudinal scaling laws is due to statistical noise, we create a null model that assumes a homogenous scaling exponent for all institutions. 
To create the null model, we fit a scaling law for each institution, keeping the residual values, $r_i$, along with their $x_i$ positions, creating a set of pairs $\{x_i,r_i\}$. The null model \textit{homogeneous scaling law}, $\beta_{\mathrm{null}}$, is the average scaling laws, $\beta_i$, across all institutions, weighted by the inverse of the standard error squared, $1/\sigma_{\beta_i}^2$. For each institution, we randomly permute the residuals $\{x_i,r_i\}\rightarrow \{x_i,r_j\}$ to create new data: $\{x_i,y_{\mathrm{null}}\} = \{x_i,x_i \beta_{\mathrm{null}}+r_j\}$. Because of random permutation of residuals, we assume the data are homoscedastic, but make no other assumptions, not even whether the residuals are normally distributed. After refitting each new set of points $\{x_i,y_{\mathrm{null}}\}$ for each institution, we expect the new null model coefficient for each institution to fluctuate around $\beta_{\mathrm{null}}$ due to noise. To see whether the distribution of null model coefficients differs from the empirically derived coefficients, we use the Kolmogorov-Smirnov test on these two distributions \cite{Massey1951}. We find almost invariably that the two distributions differ with p-value $<0.001$. 

\subsection*{Supplementary Note 4: Scaling of Output}
\begin{figure*}[thb!]
    \centering
    \includegraphics[width=0.7\textwidth]{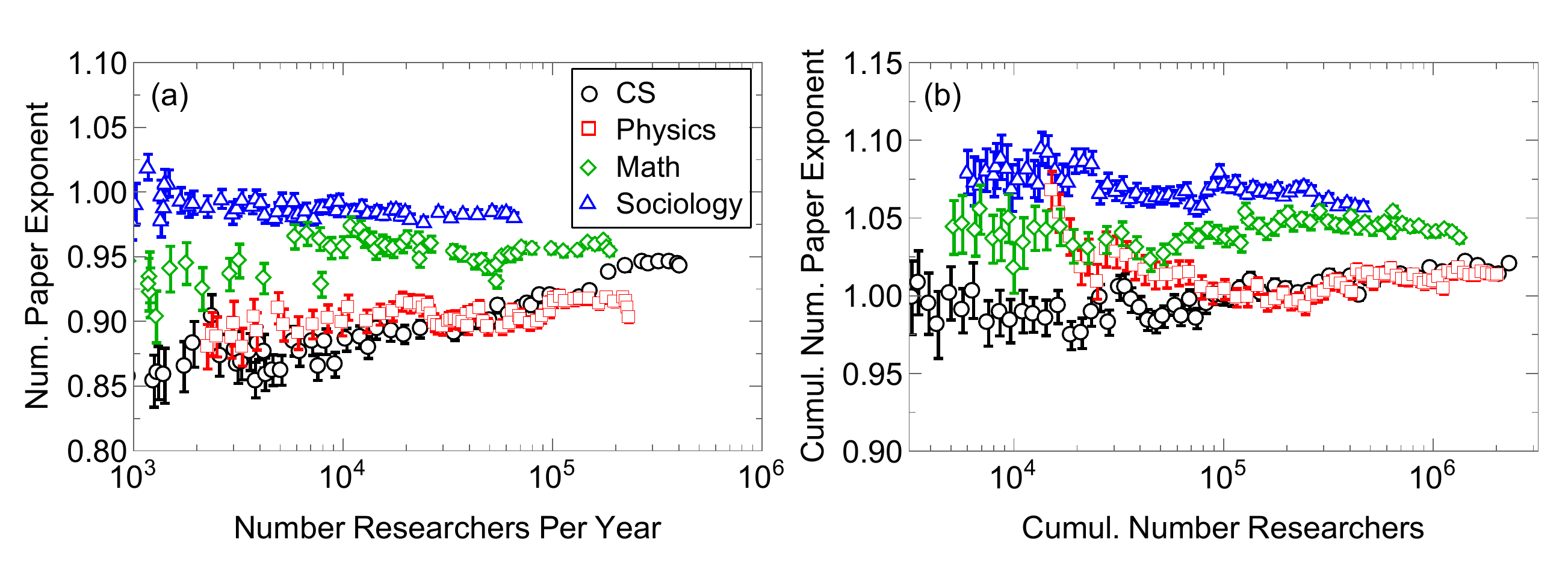}
    \caption{Scaling exponents of paper output over time from cross-sectional analysis. (a) Paper output per year and (b) cumulative paper output. We see that the number of papers scales linearly with institution size regardless of the field.}
    \label{fig:CrossInstituteProduction}
\end{figure*}

\begin{figure}[thb!]
    \centering
    \includegraphics[width=0.3\linewidth]{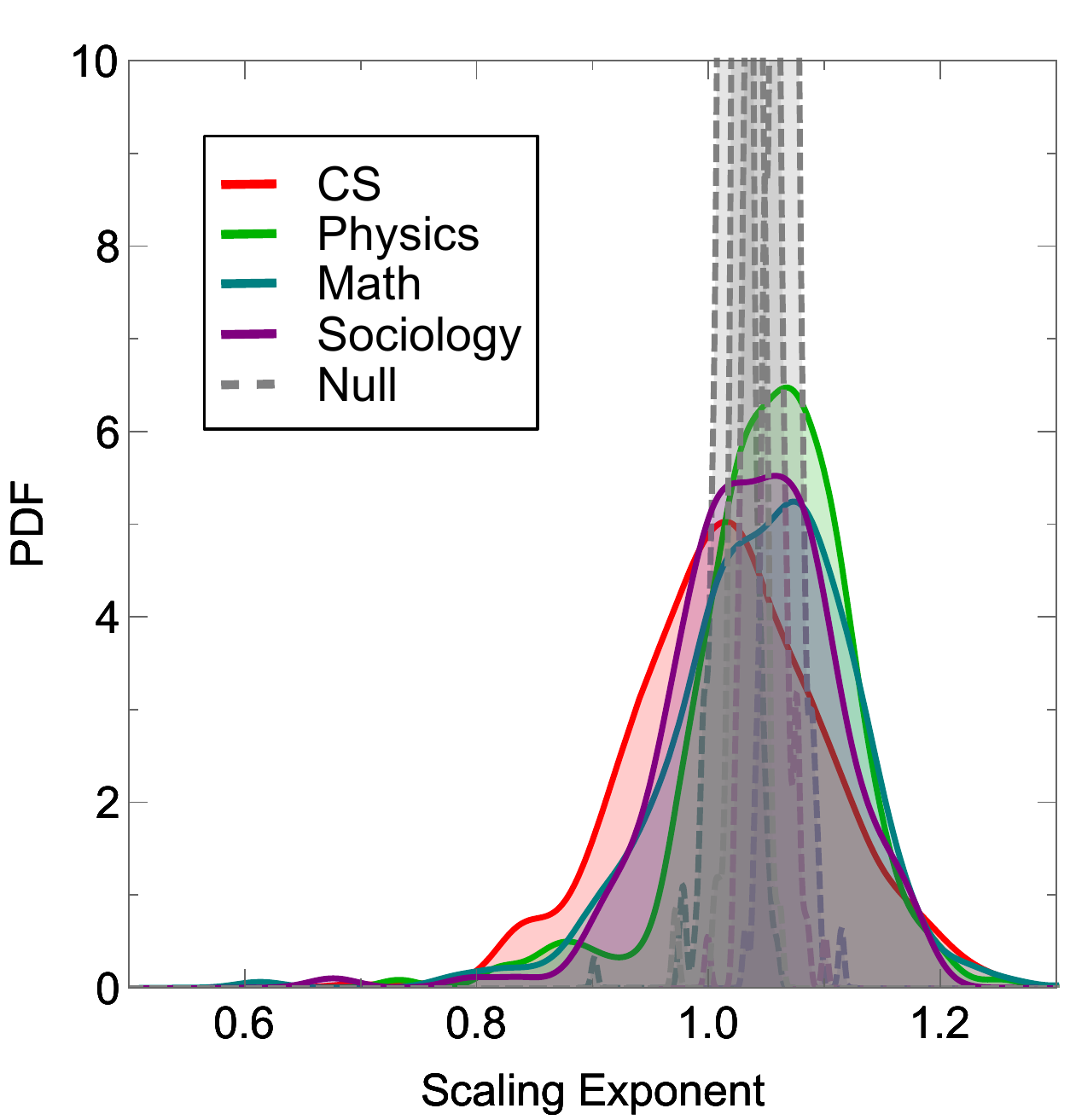}
    \caption{
    Probability distribution of longitudinal scaling exponents for institution output (number of papers) across disciplines. Gray dashed lines are the PDF of the null model in which all institutions within that field have the same scaling law. 
    \label{fig:ProductionScaling}
    }
\end{figure}

We define institution output as the cumulative number of papers written by researchers affiliated with that institution. Cross-sectional scaling laws are surprisingly stable in time, with a value of almost exactly 1.0, as shown in Fig.~\ref{fig:CrossInstituteProduction}. This means that the output per person is independent of institution size. This holds also for different disciplines, regardless of whether we look at annual output or cumulative output. 
We explore the scaling laws in longitudinal data as well in Fig.~\ref{fig:ProductionScaling}. These results also show approximately linear scaling relationships. 

Figure~\ref{fig:ProductionScaling} shows the scaling exponents of institution output versus institution size. The scaling exponents are centered around 1.0, although these scaling laws differ between institutions
. This suggests, surprisingly, that paper output per researcher is approximately independent of institution size. 
That being said, when we compare the longitudinal data to a homogeneous scaling null model, we find that institutions have a greater variance in their scaling laws than the null model  predicts. This means that some institutions create slightly more papers per person as the institution grows, while others show a reduction in output. The overall effect, however, appears to be subtle. Overall, institution size appears to affect collaborations much more than output.

\subsection*{Supplementary Note 5: Scaling of Team Size}

\begin{figure}[thb!]
    \centering
    \includegraphics[width=0.6\textwidth]{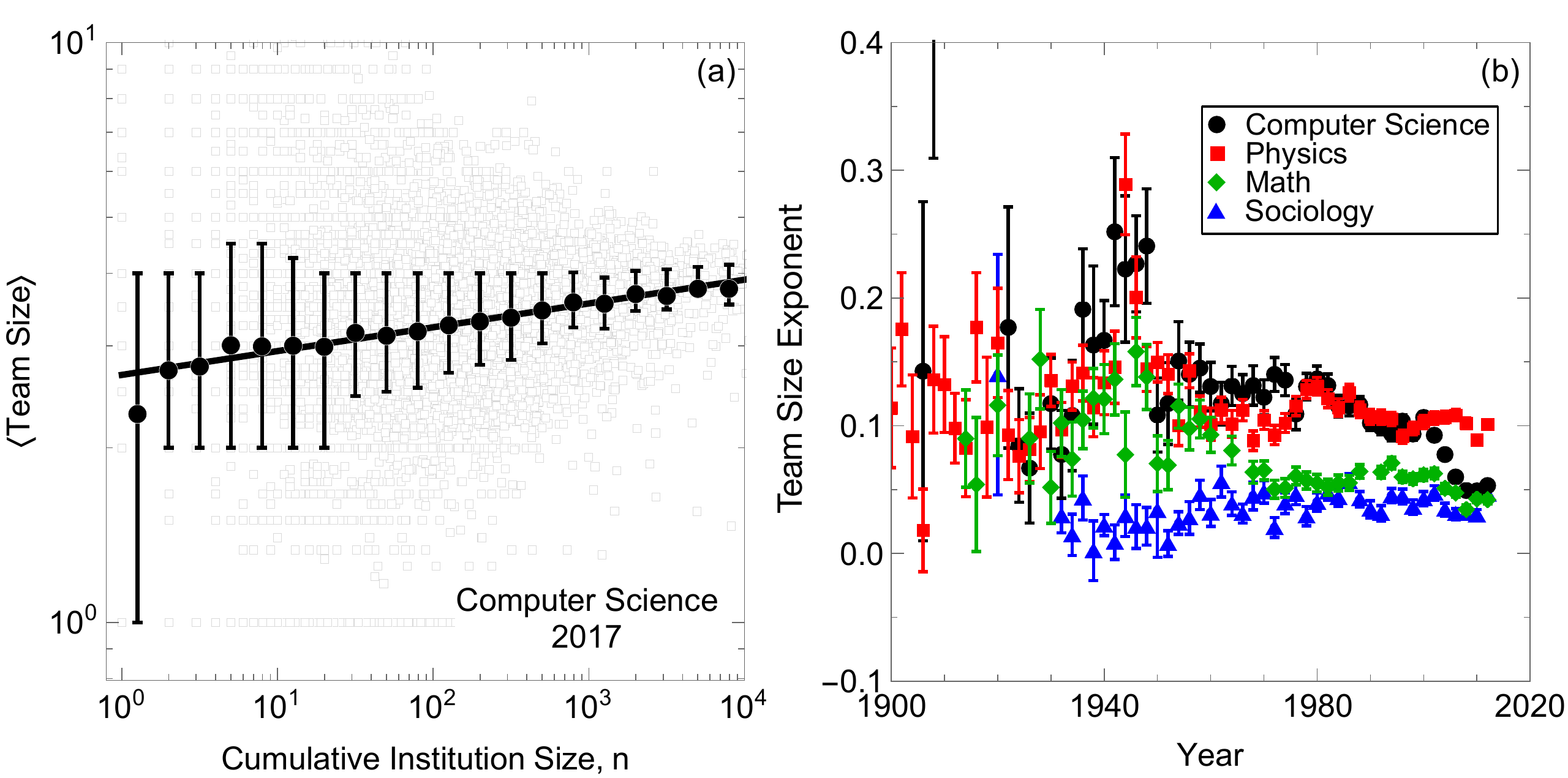}
    \caption{Cross-sectional analysis of the scaling of team size. (a) Example cross-sectional scaling for computer science in 2017, and (b) scaling exponent for each field over time. 
    }
    \label{fig:TeamCross}
\end{figure}

Team size measures the number of co-authors on a single paper. Prior work has shown that team size has grown over time, with papers produced by larger teams getting more citations compared to papers written by smaller teams~\cite{Wuchty2007}. We analyze whether institution size benefits team size via both cross-sectional and longitudinal analysis. As shown in Fig.~\ref{fig:TeamCross}, we find that team size would seem to scale positively with institution size but the scaling relations can vary significantly in time (Fig.~\ref{fig:TeamCross}b). While the scaling laws are not universal, Fig.~\ref{fig:TeamCross}a shows that the fit to a line is remarkably strong.

\begin{figure}[thb!]
    \centering
    \includegraphics[width=0.3\linewidth]{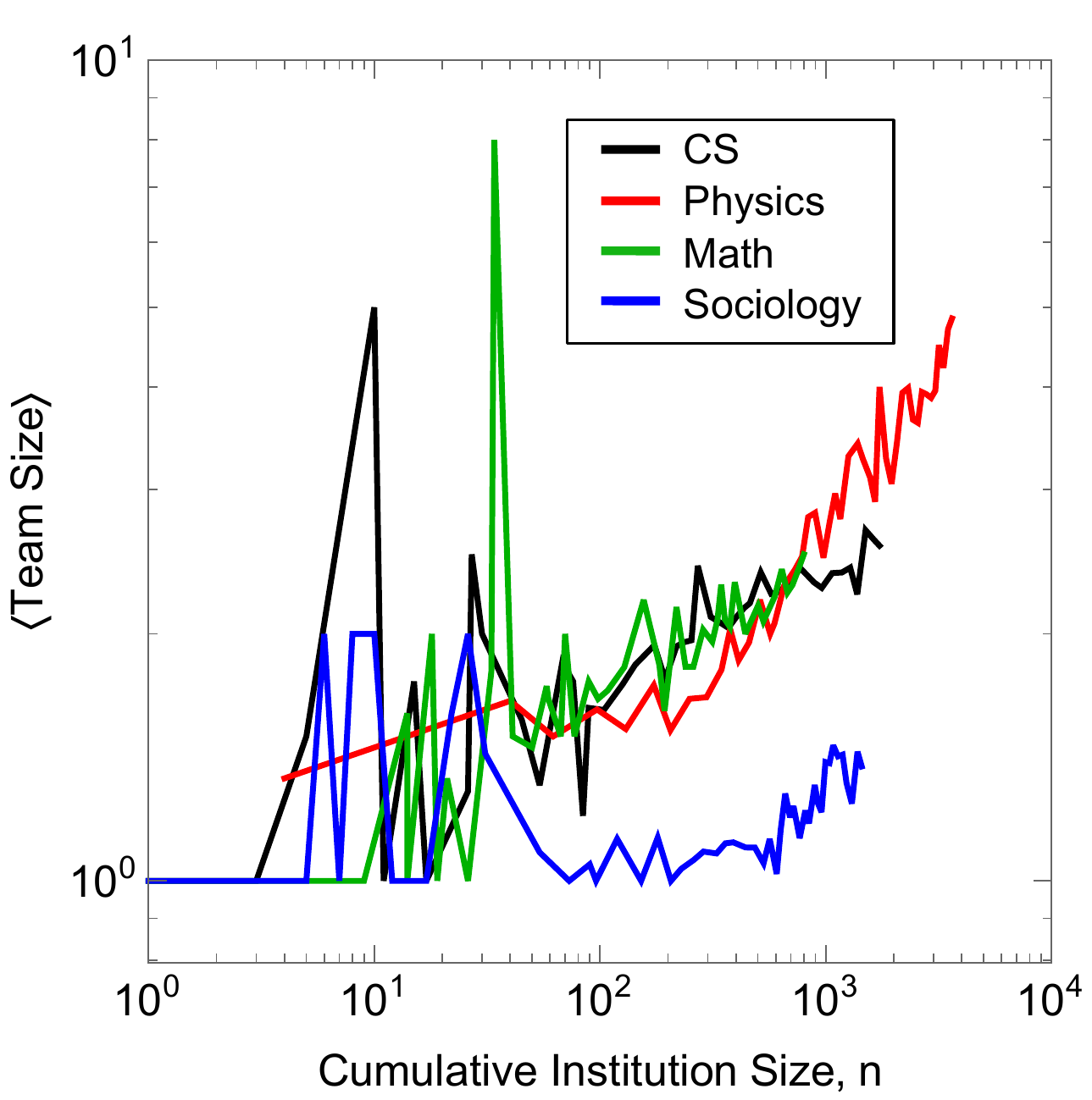}
    \caption{Longitudinal scaling examples for team size.
    }
    \label{fig:TeamLongEx}
\end{figure}

\begin{figure}[thb!]
    \centering
    \includegraphics[width=0.7\textwidth]{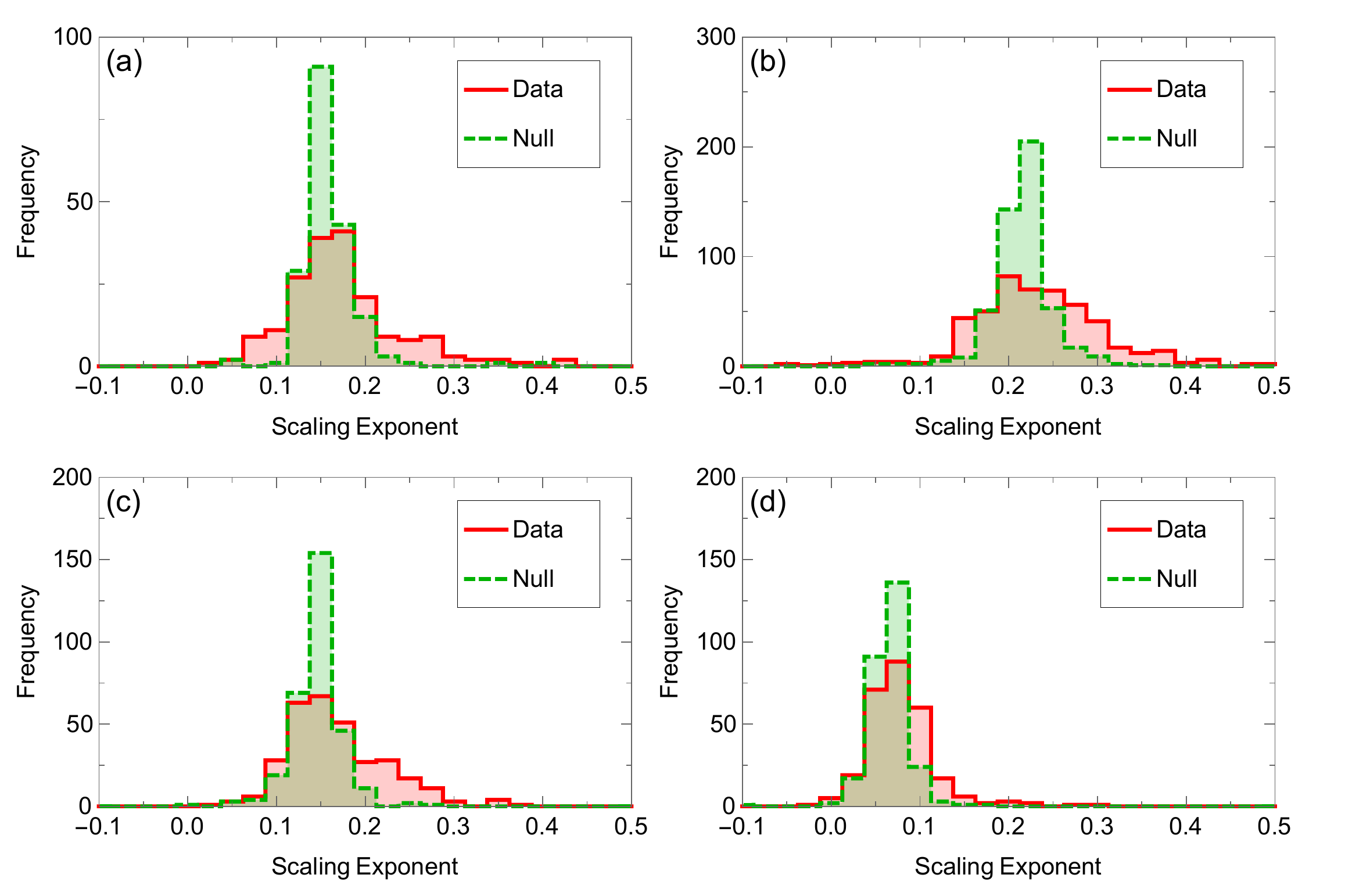}
    \caption{Histogram of longitudinal scaling exponents for team size  for (a) computer science, (b) physics, (c) math, and (d) sociology.
    }
    \label{fig:TeamLongHist}
\end{figure}

When we analyze data longitudinally, we see a more complete picture. Namely, Fig.~\ref{fig:TeamLongEx} shows that team size scales positively, but the scaling laws differ significantly between institutions. We explore this more thoroughly in Fig.~\ref{fig:TeamLongHist} where we plot a histogram of scaling exponents, whose distribution is wider than the null model (KS-test p-value $<0.001$). We see that there is both significant heterogeneity and generally larger scaling exponents than cross-sectional analysis would predict. For example, while cross-sectional analysis shows Physics has a scaling law of about 0.1, longitudinal analysis instead shows that each institution scales with an exponent of roughly 0.2.

\subsection*{Supplementary Note 6: Comparison Between Data and Simulations}
\begin{figure*}[thb!]
    \centering
    \includegraphics[width=0.6\textwidth]{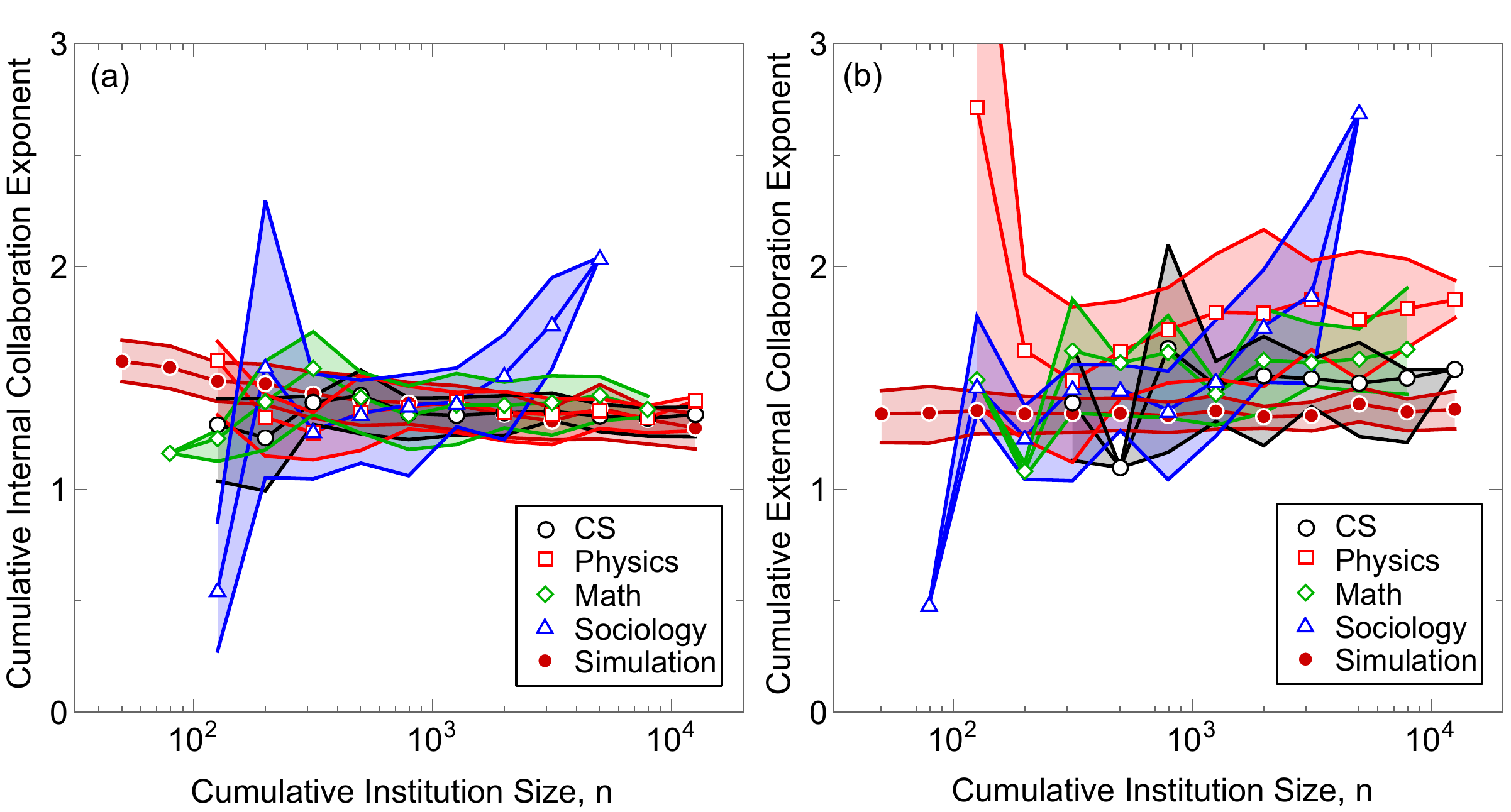}
    \caption{
    Collaboration scaling laws versus institution size. (a) Internal and (b) external collaboration exponents versus final institution size among all institutions studied. Plot points are mean scaling exponents, while shaded regions are 50\% quantiles in the distribution. In the simulation, $\rho=4$, $\nu=2$, $\mu_p=0.6$, and $\sigma_p=0.1$. 
    \label{fig:CumulBetaR2VsN}
    }
\end{figure*}

\begin{figure}[thb!]
    \centering
    \includegraphics[width=0.6\textwidth]{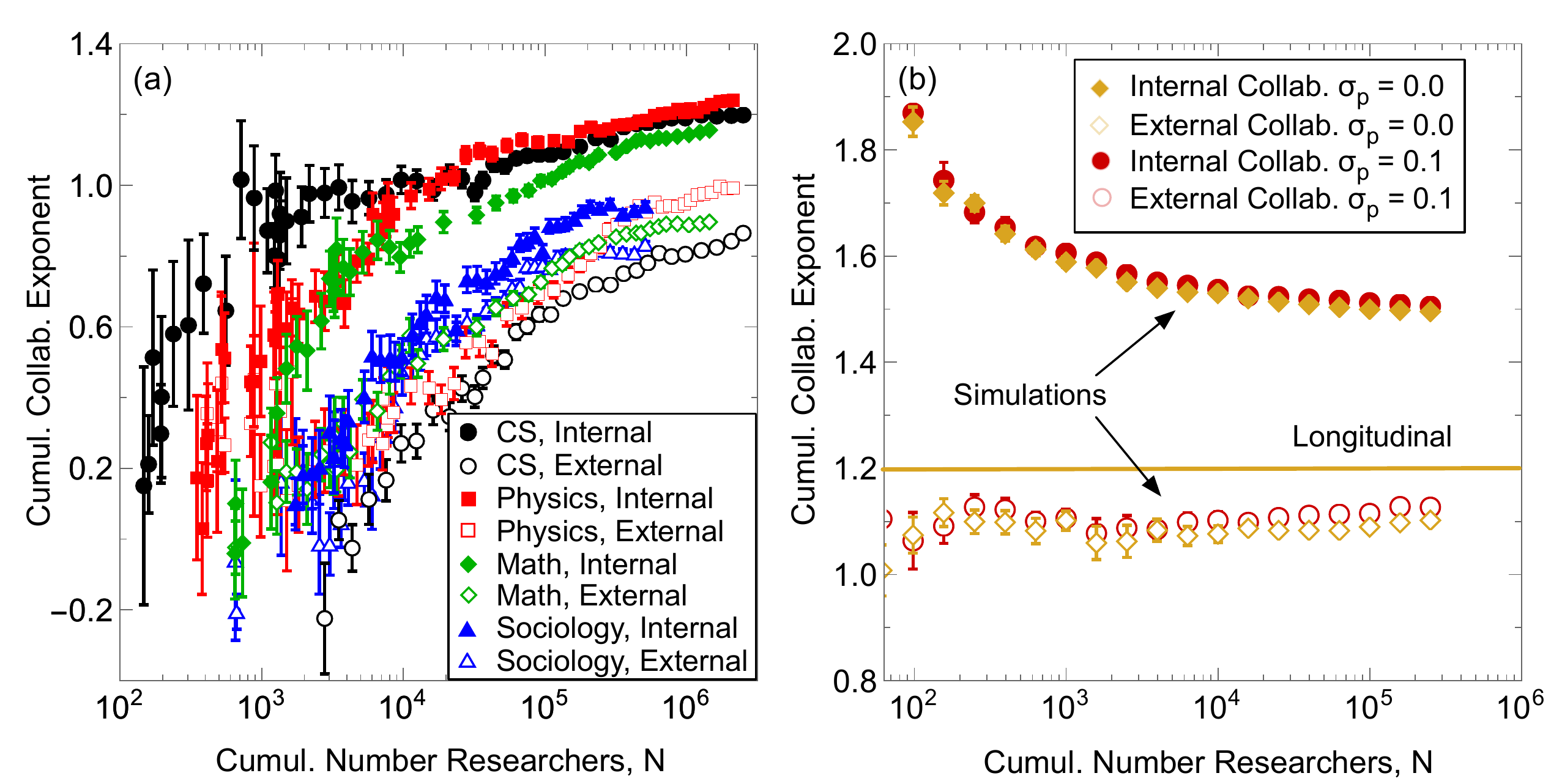}
    \caption{
    Cross-sectional analysis of the scaling of collaborations.
    (a) Internal collaboration and external collaboration scaling exponents versus the cumulative number of researchers. Error bars are standard errors. (b) Simulation cross-institution collaboration scaling versus the number of researchers. Black lines: internal collaboration, red dashed lines: external collaboration. Shaded regions are 95\% confidence intervals in the mean across different simulation realizations.}
    \label{fig:CumCrossInstituteScalingOverTime}
\end{figure}

For the rest of the SI, we will just discuss cumulative results. Figure~\ref{fig:CumCrossInstituteScalingOverTime}a is similar to the main text by showing that cumulative cross-scaling exponents vary, but here the x-axis is the total number of researchers who have authored a paper up until that date. This slightly unusual x-axis allows us to compare these results to cross-sectional scaling in simulations shown in Fig.~\ref{fig:CumCrossInstituteScalingOverTime}b. Parameters in some of these simulations are the same as the main text, with $\nu=2$, $\rho=4$, $\mu_p=0.6$, and $\sigma_p=0.1$. For these parameters, we discover that, much like in the data, internal scaling laws are higher than external scaling laws, even though, for each institution, both should be centered around the horizontal lines labeled ``longitudinal'' (which corresponds to the mean values in the longitudinal analysis. We also notice that, like the data, the exponents vary as a function of the total number of researchers. These simulations do not just allow us to reproduce results, however, but we can make contrapositive hypotheses. For example, what would the statistics look like if there was no statistical variation in the longitudinal scaling laws? To better understand this, we let $\sigma_p=0.0$ in Fig.~\ref{fig:CumCrossInstituteScalingOverTime}b, and discover that the results are quantitatively almost exactly the same. If institutions had the exact same scaling laws and the exact same constant coefficients, then the cross-sectional and longitudinal scaling laws would be the same. These discrepancies point to either finite size effects or different constant coefficients are dominant factors in explaining why cross-sectional scaling and longitudinal scaling laws differ and vary in time, at least in simulations. We hypothesize similar effects in empirical data as well, although there are qualitative differences in data, such as scaling laws increasing rather than decreasing which point to limitations of the simulations. We also show in Fig.~\ref{fig:CumulBetaR2VsN} that internal and external scaling laws sare intrinsic, and do not vary significantly with institution size. Moreover, there is significant variance in these scaling laws that our simulation can capture.

\begin{figure}[thb!]
    \centering
    \includegraphics[width=0.6\textwidth]{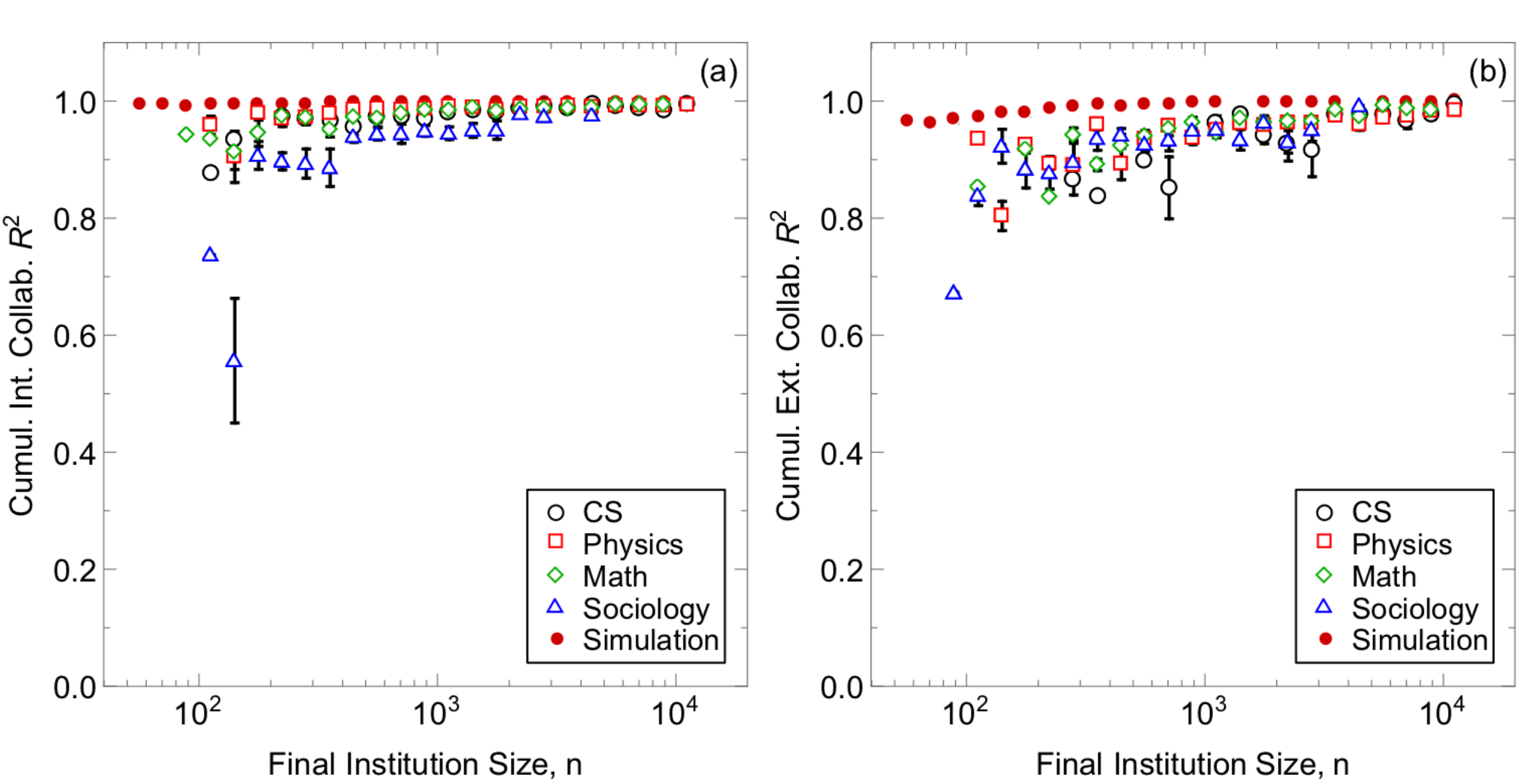}
    \caption{Quality of scaling law fits of (a) internal and (b) external collaborations. The $R^2$ metric of log-log fits averaged for all institutions of size $n$ in 2017 quickly approaches 1.0 as $n$ grows. Error bars are standard errors.}    \label{fig:logfit-CumCrossInstituteScalingOverTime}
\end{figure}

Focusing on longitudinal scaling, however, we observe in Fig.~\ref{fig:logfit-CumCrossInstituteScalingOverTime} that the data and simulations are both well-characterized by linear relations in log-log space. Namely, the figure shows that a linear fit of log(collaborations) versus log(institution size) have an $R^2$ value of nearly 1 for each institute. If their size as of 2017 is large, then $R^2$ is even closer to 1, in agreement with what we should expect in the thermodynamic limit (where finite size effects are negligible). The data is therefore well-characterized as a power law, but these power law values vary between institutions, as shown in Fig. 3 of the main text. 

\begin{figure}[thb!]
    \centering
    \includegraphics[width=0.6\textwidth]{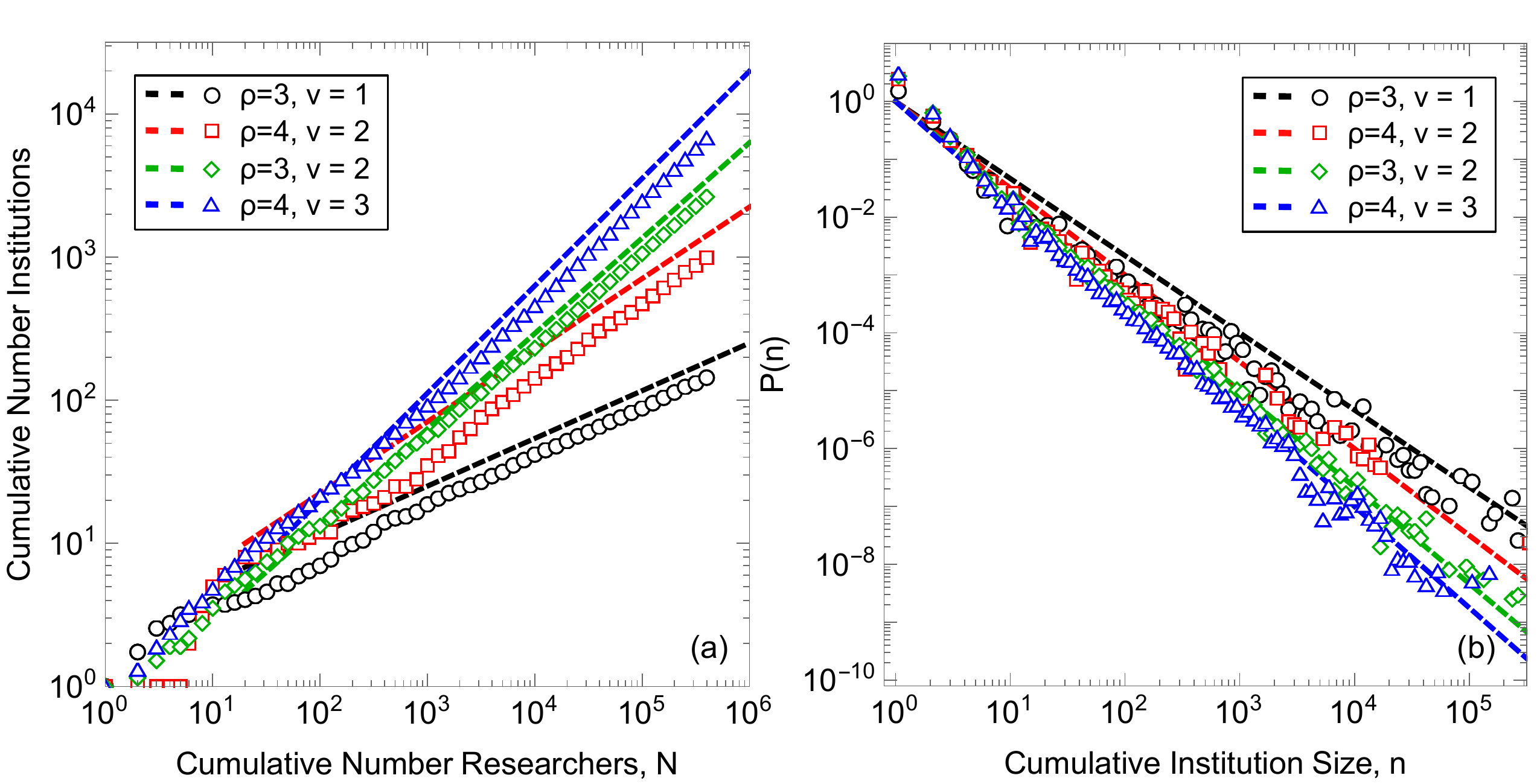}
    \caption{Simulation of (a) Heaps law and (b) Zipf's for various values of $\rho$ and $\nu$. Dashed lines are theoretical scaling exponents.
    }
    \label{fig:SimHeapsZipfs}
\end{figure}

Theory surrounding the Polya's urn portion of our model is discussed in detail in previous work \cite{Tria2014}. Nonetheless, we check the robustness of this theory in Fig.~\ref{fig:SimHeapsZipfs}. We find excellent agreement between the theory and simulation, demonstrating that, even for finite sizes, the theory they developed accurately explains the simulation patterns.

\subsection*{Supplementary Note 7: Robustness Check of Simulations}

\begin{figure}[thb!]
    \centering
    \includegraphics[width=0.7\textwidth]{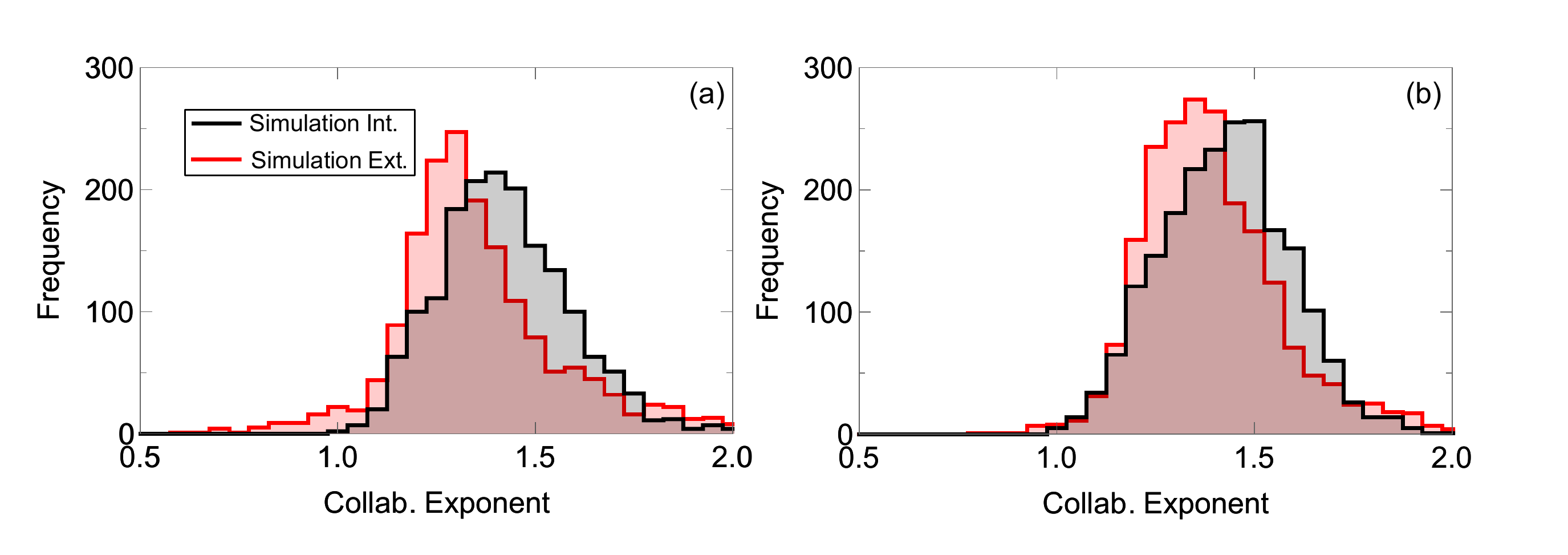}
    \caption{Internal and external longitudinal collaboration exponents for alternative simulation models. (a) Internal and external exponents for simulations with $\lambda=1$ Poisson distributed numbers of initial collaborators (on average one internal collaborator, and one external collaborator). (b) The same histograms for the current simulation with exactly one internal and one external collaborator.}
    \label{fig:PoissonVsDeterministicLong}
\end{figure}

\begin{figure}[thb!]
    \centering
    \includegraphics[width=0.6\textwidth]{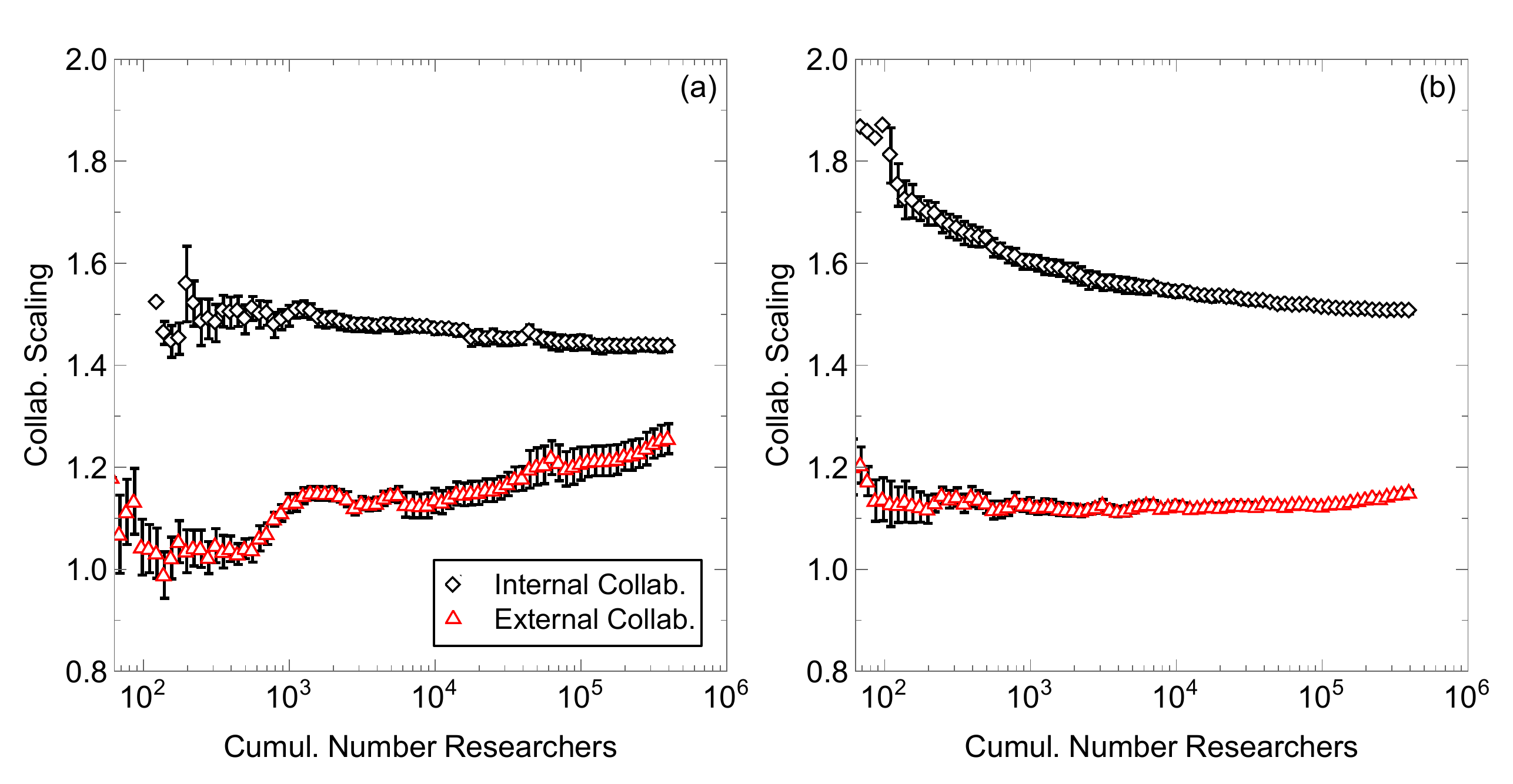}
    \caption{Internal and external cross-sectional collaboration exponents for alternative simulation models. (a) Internal and external exponents versus the cumulative number of researchers for simulations with $\lambda=1$ Poisson distributed numbers of initial collaborators (on average one internal collaborator, and one external collaborator). (b) The same figure for the current simulation with exactly one internal and one external collaborator.}
    \label{fig:PoissonVsDeterministicCross}
\end{figure}

We might wonder whether our model is sensitive to stochastic variations in how the model behaves. For example, we might ask whether changing the number of initial collaborators from 1 to a range of values will affect results. To this end, we made an additional model in which the number of initial internal and external collaborators was Poisson distributed, with $\lambda=1$ (i.e., on average one internal and one external collaborator). This will not affect the institution formation, but it might affect the institution growth, e.g., the longitudinal collaboration scaling exponents. Importantly, Bhat et al. and Lambiotte et al. shows that number of links over time are not self-averaging \cite{Lambiotte2016,Bhat2016}, therefore initial conditions greatly affect the final number of links. Figures~\ref{fig:PoissonVsDeterministicLong} \& \ref{fig:PoissonVsDeterministicCross} show our results. In Fig.~\ref{fig:PoissonVsDeterministicLong}, we find that, while there are slightly more outliers in the scaling exponent distribution, results are quantitatively very similar. In Fig.~\ref{fig:PoissonVsDeterministicCross}a we find that $\lambda=1$ external collaboration cross-sectional scaling exponents increase with the cumulative number of researchers, more alike to what we see in empirical data (Fig.~2 main text), and the internal collaboration exponents are stationary. In Fig.~\ref{fig:PoissonVsDeterministicCross}b, however, we find that the external collaboration exponents are mostly stationary in the original form of the model, while internal collaboration exponents decrease with the cumulative number of researchers.


\section*{Acknowledgments}
Research was funded by in part by DARPA under contract \#W911NF1920271 and by the USC Annenberg Fellowship. Data, code used for data analysis, code for simulations is available in the following repository: https://github.com/ZagHe568/institution\_scaling.

\end{document}